\documentclass[sigconf, anonymous=false, nonacm=true, pbalance=true]{acmart}
\usepackage{multirow}
\usepackage[most]{tcolorbox}
\usepackage{booktabs}
\usepackage{tabularx}
\usepackage{longtable}
\usepackage{array}
\usepackage{pifont}
\usepackage{makecell}
\usepackage{hyperref}

\newtcbtheorem{Finding}{\bfseries Finding}{enhanced,drop shadow={black!50!white},
  coltitle=black,
  top=0.3in,
  attach boxed title to top right=
  {xshift=0em,yshift=-\tcboxedtitleheight/2},
  boxed title style={size=small,colback=darkgray}
}{finding}

\usepackage{tcolorbox}
\definecolor{verylightgrey}{gray}{0.9}
\newtcolorbox{takeaway}{
  colback=verylightgrey,
  colframe=verylightgrey,
  sharp corners,
  boxrule=0mm,
  boxsep=0mm,
  left=1mm,
  right=1mm,
  top=1mm,
  bottom=1mm
}

\newcounter{myfindingscounter}
\newcommand\takeawaytitle[1]{\textbf{Finding \refstepcounter{myfindingscounter}\themyfindingscounter\label{#1}:}}

\usepackage{xspace}

\newcommand{\mypara}[1]{\vspace{4pt}\noindent{\textbf{{#1}}\xspace}}

\AtBeginDocument{%
  }

\setcopyright{acmlicensed}
\copyrightyear{2018}
\acmYear{2018}
\acmDOI{XXXXXXX.XXXXXXX}

\acmISBN{978-1-4503-XXXX-X/18/06}

\acmSubmissionID{2085}

\newcommand{\supplementarysection}{%
  \setcounter{figure}{0}% Reset figure counter
  \let\oldthefigure\thefigure% Capture figure numbering scheme
  \renewcommand{\thefigure}{S\oldthefigure}% Prefix figure number with S
  \setcounter{table}{0}% Reset figure counter
  \let\oldthetable\thetable%
  \renewcommand{\thetable}{S\oldthetable}%
}

\begin{document}

\title{Digital Gatekeeping: An Audit of Search Engine Results shows tailoring of queries on the Israel-Palestine Conflict}

\author{Íris Damião}
\orcid{0009-0005-4931-2376}
\affiliation{
  \institution{LIP - Laboratory for Instrumentation and Particle Physics \&  Instituto Superior Técnico - University of Lisbon}
  \city{Lisbon}
  \country{Portugal}}
\email{irisdamiao@lip.pt}

\author{José M. Reis}
\orcid{0000-0002-8055-0170}
\affiliation{
 \institution{LIP, Laboratory for Instrumentation and Particle Physics*}
 \city{Lisbon}
 \country{Portugal}}
 \thanks{* current affiliation: Portuguese National Cybersecurity Centre, Lisbon, Portugal.}
 
\author{Paulo Almeida}
\orcid{0000-0002-9279-2353}
\affiliation{
 \institution{LIP, Laboratory for Instrumentation and Particle Physics}
 \city{Lisbon}
 \country{Portugal}}

\author{Nuno Santos}
\orcid{0000-0001-9938-0653}
\affiliation{%
  \institution{INESC-ID \& Instituto Superior Técnico, University of Lisbon}
  \city{Lisbon}
  \country{Portugal}
}

\author{Joana Gonçalves-Sá}
\orcid{0000-0001-6654-2126}
\affiliation{
  \institution{LIP, Laboratory for Instrumentation and Particle Physics \& NOVA LINCS, FCT NOVA University}
  \city{Lisbon \& Caparica}
  \country{Portugal}}
\email{joanagsa@lip.pt}

\renewcommand{\shortauthors}{Damião et al.}

\begin{abstract}

Search engines, often viewed as reliable gateways to information, tailor search results using customization algorithms based on user preferences, location, and more. While this can be useful for routine queries, it raises concerns when the topics are sensitive or contentious, possibly limiting exposure to diverse viewpoints and increasing polarization.

To examine the extent of this tailoring, we focused on the Israel-Palestine conflict and developed a privacy-protecting tool to audit the behavior of three search engines: DuckDuckGo, Google and Yahoo. Our study focused on two main questions: (1) How do search results for the same query about the conflict vary among different users? and (2) Are these results influenced by the user's location and browsing history?

Our findings revealed significant customization based on location and browsing preferences, unlike previous studies that found only mild personalization for general topics. Moreover, queries related to the conflict were more customized than unrelated queries, and the results were not neutral concerning the conflict's portrayal.

\end{abstract}

%% ATTENTION - CHANGE BEFORE SUBMISSION


% %% A "teaser" image appears between the author and affiliation
% %% information and the body of the document, and typically spans the


\maketitle

\section{Introduction}

Access to high-quality information is a central tenet of democratic regimes, allowing citizens to make informed decisions. In recent decades, the way people seek information has shifted from traditional to online media \cite{pew_news_2006, news_access, ReutersInstitute2024}, with search engines playing a crucial role as gate keepers. Google, for instance, processes around 8.5 billion daily queries \cite{google_queries, google_maximize_access}, highlighting its influence on global information consumption.

While this shift has greatly enhanced the speed and convenience of accessing information, it also raises concerns about how search results shape users' worldviews. Studies show that individuals searching for news place strong trust in search engine results \cite{misplaced_trust, google_trust, trust_survey_2012, trust_survey_2017, trust_survey_2020}, often seeing these results as the most relevant \cite{EuropeanCommission2016}. Additionally, search engines rely on customization algorithms, which tailor content results based on user preferences, location, and other factors \cite{duckduckgo_privacy, google_terms, Yahoo2024, inconsistent_search_results, personalization_web_search}. While this facilitates routine searches, such as finding a restaurant, it becomes problematic for sensitive topics, like political news, where customized results can introduce bias and reinforce ``filter bubbles'' \cite{eli-pariser, personalization_web_search}. Given the opaque nature of search engine algorithms, which tend to monetize clicks, it is difficult to determine how much of this tailoring affects controversial topics compared to less sensitive ones. If differences exist, greater transparency is needed to explain how and why search engines suggest user-specific results.

Prior research has examined driving factors of customization, such as location, browsing history, and user accounts~\cite{inconsistent_search_results, personalization_web_search, political_personalization}, but these studies are often conducted in isolation, analyzing a single feature at a time, and predominantly examining Google. Much of this research has also focused either on broad, consensual topics or on political events such as elections, leaving a gap in understanding how search engines handle deeply polarized and contentious issues.

In this paper, we aim to address this gap by presenting the first empirical study on the effects of search engine customization on global controversial topics. We focus on the contemporary Israel-Palestine conflict and analyse the search results offered by three popular search engines---DuckDuckGo, Google, and Yahoo. We chose to study this conflict not only because of the seriousness of this polarizing global affair but also because it involves stakeholders across different political, geographical, cultural, and linguistic boundaries. This diversity offers a particularly rich context to examine how factors like user location, language, and browser history influence search engine customization. We explore three main research questions: \textit{1)} Do search engine results show higher levels of customization for more polarizing queries compared to non-sensitive topics, and, if so, what factors drive this tailoring? \textit{2)} How different in type and content are the URLs provided by the search engines for conflict-related searches when compared to general, non-sensitive topics? \textit{3)} Do the suggested contents vary according to geographical location and, if so, do they carry particular conflict-related leanings?
%Do the personalization factors used by the search engines influence the leanings of the results concerning the conflict? 

Performing this study presented several non-trivial challenges. First, search engines operate mostly as black boxes, making it difficult to systematically examine how their algorithms function. Second, it is essential to mimick realistic user behavior across different regions, languages, and browsing habits without introducing bias. Third, it is neither trivial to distinguish between neutral and politically charged queries to draw meaningful conclusions, and define reliable metrics to measure not only the variations in search engine results. This paper addresses these challenges and makes the following key contributions:

\mypara{1. Crawling method with incremental profiling:} We developed an automated crawling approach to perform online searches using software bots. These bots were progressively made more ``human-like'' by expanding their profiles: \textit{Type 1} bots varied only by location, \textit{Type 2} bots included different browser languages, and \textit{Type 3} bots incorporated an initial online browsing period to collect cookies. We deployed these bots in various countries and directed them to the search engines to perform identical queries. These queries included both conflict-specific topics (e.g., ``Hamas'') and general, non-specific topics (e.g., ``how to tie a tie''). In this way, we maintained control over each feature's contribution to customization while avoiding the use of private or sensitive user data.

\mypara{2. Metrics for measuring customization:} After deploying the bots, we analyzed whether the results varied and whether these variations were reflected in differences in content or in conflict-related bias. To do this end, we adopted a metric system based on Rank-Biased Overlap (RBO)~\cite{rank_biased_precision_measurement}, which emphasizes top results, and employed additional methods to assess qualitative differences (details below). This framework enables us to study how individual features combine to influence search engine results and serves as a tool to evaluate accuracy and potential biases beyond the Israel-Palestine case study.

\mypara{3. Significant findings:} Our study reveals a higher level of search engine customization than previously reported, even for sensitive topics like electoral events \cite{haim2017burst, krafft2019search_engine_manipulation, partisan_audience_bias}. Results were particularly pronounced for conflict-related queries, and when comparing across locations. Moreover, and despite claims from some search engines regarding limited or no personalization based on user data, browsing history influences the results displayed. In particular, Google and DuckDuckGo, which claim to limit personalization based on browsing history, showed significant differences in the results delivered, while Yahoo, which openly acknowledges more personalization, showed the least.

\vspace{-0.1cm}
\section{Background and Related Work}
Search engines apply various criteria to determine the information presented online, such as keyword matching, site authority, context, location, and language \cite{google_how_search_works, moz_how_search_engines_operate}. These algorithms choices can help shape individual views on particular topics, and studies have shown that even small changes in rank biased towards a political candidate can shift the voting preferences of undecided voters by 20\% \cite{epstein2015search_engine_manipulation}. As the factors used to tailor results to a specific user, remain uncertain, despite their potentially large impacts \cite{eli-pariser}, it is fundamental to develop independent tools to explore these choices. In fact, it is often impossible to distinguish if
differences are due to personalization, here defined as \textit{search results curated particularly for a user in question, considering the user's tastes and specific profile} or to customization, which, may result from differences in less personal features, such as location or language.

This work focuses on how different features can impact the order, the content, and the leaning of search-engine results, especially when a user is looking for geographical and politically sensitive information, as is the case of the Israel-Palestine conflict. 
%Particularly, when small changes can have big impacts of users perception \cite{epstein2015search_engine_manipulation}. 

%(so that when searching for a restaurant called "Berlin" in California, we do not get directed to a restaurant in Germany).
%Therefore, thorough out the paper, we will refer to personalization only when search engine differences due to other factors than our bot direct profile
%In this section, we review search engines' policies regarding results personalization and customization, and some past efforts to audit their recommendation systems.

\subsection{Search Engine Customization}

%Previous research has examined various aspects of search engine personalization, such as location, user accounts, and browsing history \cite{personalization_web_search, inconsistent_search_results}, but there is ongoing debate about the extent of personalization in search engines \cite{krafft2019search_engine_manipulation, partisan_audience_bias, bhaim2017burst} and the factors that drive it \cite{political_personalization, auditing_personalization_and_composition}. While these studies offer valuable insights, they are often conducted in isolation (a single user behavioral feature being analyzed), focusing predominantly on Google \cite{inconsistent_search_results, what_did_you_see, partisan_audience_bias, i_vote_for, auditing_personalization_and_composition}, and typically examine either broad, consensual topics \cite{inconsistent_search_results, personalization_web_search} or exclusively political issues surrounding electoral events \cite{auditing_personalization_and_composition, political_personalization}.

In this section, we briefly review search engine policies regarding result personalization and customization, covering the search engines we study in this work. Table \ref{engines_policies} summarizes these policies.

\textbf{DuckDuckGo} is known for its privacy-concerned model, explicitly stating it does not log IP addresses, browsing history, or search history to determine results. However, while it does not use exact user locations, it provides geographical results through proxies~\cite{duckduckgo_privacy}.

\textbf{Google} is the most widely used search engine, and claims that results may differ from user to user based on ``time, context and personalization'' \cite{Google2023}. Time differences arise from data being updated at various speeds across data centers. Context includes aspects such as location, language, device type (screen, mobile or desktop), and related results, which may appear if a user engages with specific content on the results page. Personalized results - specifically tailored to an individual user profile — are claimed to be tied to a Google Account.

\textbf{Yahoo} offers a search engine, but its primary business focus lies on email and news services, with these being particularly popular \cite{similarweb}. It openly claims to use past browsing history and search queries across all its recommendation systems, including email and news \cite{Yahoo2024}. Additionally, Yahoo email users can receive personalized search content based on their activity whithin these cross-platform services. % Yahoo is an interesting case because, although it offers a search engine, its primary business focus lies in its email and news services, with these being particularly popular [REF]. According to Yahoo's \textit{Information Collection \& Use Practices page} [REF], information such as past browsing history and search queries is used across all of its recommendation systems. Notably, due to Yahoo's specific services, some users may also receive personalized results from other sources, such as Yahoo Mail. These private results, however, are only displayed when the user is logged in. 

\begin{table}[h]
\centering
\caption{Customization Features Disclosed by Search Engines' Terms of Service}
\begin{tabular}{>{\raggedright\arraybackslash}p{4cm}|>{\centering\arraybackslash}p{1.1cm}|>{\centering\arraybackslash}p{0.9cm}|>{\centering\arraybackslash}p{0.9cm}}
\textbf{Feature} & \textbf{DuckGo} & \textbf{Google} & \textbf{Yahoo} \\ \midrule
Location & \textasciitilde proxy & \ding{51} & \ding{51} \\ \midrule
Language & \ding{51} & \ding{51} & \ding{51} \\ \midrule
Device Type & \ding{109} & \ding{51} & \ding{51} \\ \midrule
Search History (\small{within platform}) & \ding{55} & \ding{108} & \ding{51} \\ \midrule
Browsing History & \ding{55} & \ding{55} & \ding{51} \\ \midrule
Previous Content Engagement & \ding{55} & \ding{108} & \ding{55} \\ \midrule
Private Data (\small{e.g., Email}) & \ding{55} & \ding{55} & \ding{108} \\ \bottomrule 
\end{tabular}
\caption*{\ding{51} - feature impacts search results\\
\ding{55} - feature does not impact search results\\
\ding{108} - feature impacts results when user is logged in\\
\ding{109} - policy is unclear on whether feature impacts results}
\label{engines_policies}
\end{table}

% \begin{table*}[ht]
% \centering
% \begin{tabular}{|>{\raggedright\arraybackslash}p{2cm}|>{\centering\arraybackslash}p{1.5cm}|>{\centering\arraybackslash}p{1.5cm}|>{\centering\arraybackslash}p{1.5cm}|>{\centering\arraybackslash}p{1.5cm}|>{\centering\arraybackslash}p{1.5cm}|>{\centering\arraybackslash}p{2.0cm}|>{\centering\arraybackslash}p{3cm}|}
% \hline
% \textbf{Search Engine} & \textbf{Location} & \textbf{Language} & \textbf{Device Type} & \textbf{Search History} & \textbf{Browsing History} & \textbf{Previous Content Engagement} & \textbf{Private Data (e.g., Email)} \\ \hline
% \textbf{Google} & \ding{51} & \ding{51} & \ding{51}  & \ding{108} (logged-in only)  & \ding{108} (logged-in only)  & \ding{108} (logged-in only)   & \ding{55}  \\ \hline
% \textbf{Bing}  & \ding{51} & \ding{51} & \ding{51}  & \ding{51} & \ding{109}  & \ding{109}  & \ding{55} \\ \hline
% \textbf{DuckDuckGo} & \ding{108} (proxy) & \ding{51}  & \ding{109}  & \ding{55} & \ding{55} & \ding{55} & \ding{55} \\ \hline
% \textbf{Yahoo}  & \ding{51}  & \ding{51} & \ding{51} & \ding{51} & \ding{51}  & \ding{55} & \ding{108} (logged-in only)  \\ \hline
% \end{tabular}
% \caption{Comparison of Personalization Features Across Search Engines}
% \label{engines_policies}
% \end{table*}
\subsection{Search Engine Audits}

Given the proprietary nature of search engine algorithms, researchers have relied on external audits to evaluate the extent and possible negative consequences of search engine customization.
%Since Eli Pariser's publication of the book "The Filter Bubble: What the Internet is Hiding from You" \cite{eli-pariser} raised concerns about search engine personalization, the scientific community has approached the issue using various approaches and methodologies. 
In 2013, Hannak et al.~\cite{personalization_web_search} introduced a methodology based on comparing the results of a ``real'' person with those of a ``fresh'' profile for the same query. They showed that ``location'' and ``being logged-in'' were the most important factors to induce search result differences. This work only audited Google but was later expanded to include Bing, which, on average, showed even more personalized results~\cite{measuring_personalization_2}.

Due to its importance, most past studies focused on political customization in the context of electoral processes, and almost solely audit Google~\cite{political_personalization, auditing_personalization_and_composition, partisan_audience_bias, krafft2019search_engine_manipulation, i_vote_for}. These often examine the effects of different factors in isolation (typically either location or browsing history), and have produced mixed findings. While some earlier research found no significant personalization in search results based on users' political profiles \cite{krafft2019search_engine_manipulation, partisan_audience_bias, haim2017burst}, more recent studies suggest that a user's political leaning can influence the political bias of results, especially in top news listings~\cite{political_personalization, auditing_personalization_and_composition}. Moreover, location has been shown to have a stronger influence on local queries than on political ones~\cite{inconsistent_search_results}. Importantly, and unlike our study, this literature primarily compares URLs, and only to a smaller extent, it investigates differences in the content that these URLs direct to. Important exceptions are the work of Le et al.~\cite{political_personalization}, measuring the leaning of search engine results page domains, but not of individual URLs, of Robertson et al.\cite{auditing_personalization_and_composition}, that groups search results in categories, and Hu et al.~\cite{10.1145/3308558.3313654}, that scores partisan bias of results' snippets and the corresponding web pages.

In contrast, our work aims to explore not only the impact of individual features on a single search engine, but how the incremental combination of several significant factors — those highlighted by previous studies and publicly acknowledged by search engines (Table \ref{engines_policies}) — affects search outcomes across platforms. Even if these differences are not strictly due to personalization, if they create noticeable variations in ranking and content, they may shape users' perceptions, on a timely and polarizing topic.

\begin{figure*}[t]  \includegraphics[width=0.95\textwidth]{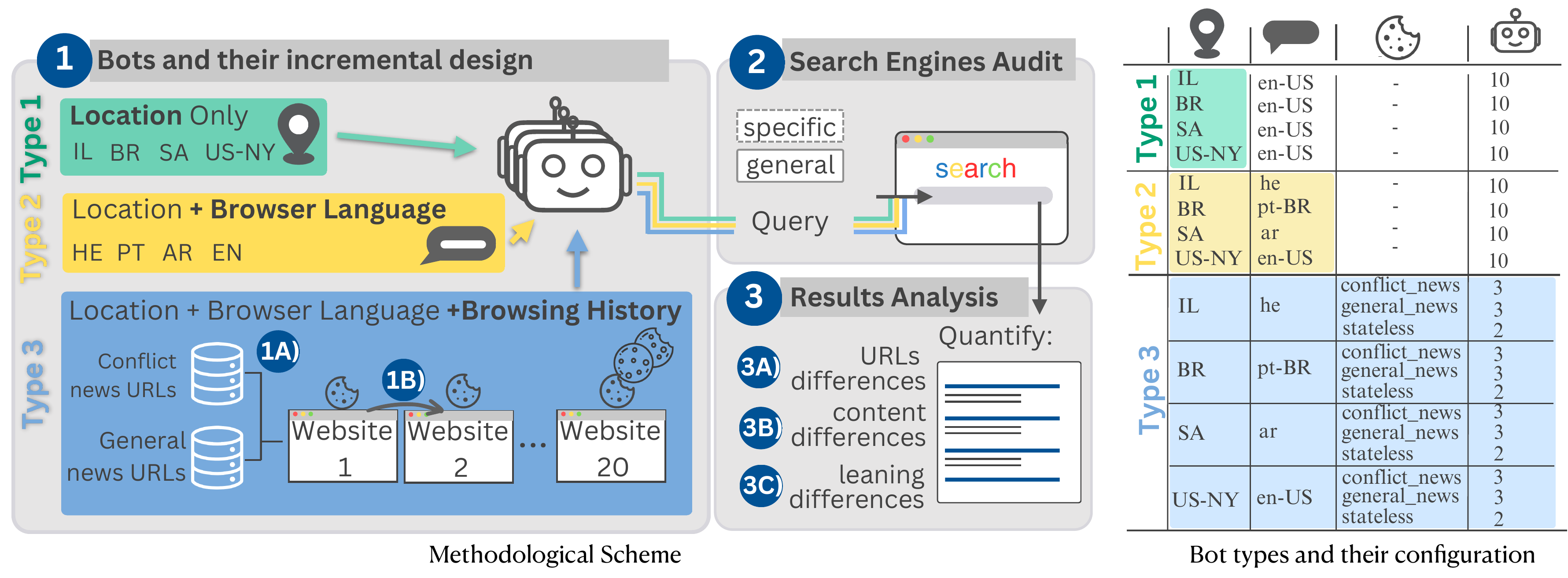}
  \caption{Left side: Methodological scheme. Three types of bots were deployed (1): Type 1 differ in location, Type 2 in location and language, Type 3 in location, language and browsing history, having collected cookies. These bots are directed to search engines and make general or conflict specific queries (2). The results are analysed according to URL and differences in content and leaning (3). Right side. Table summarizes the features of the three types of bots. Order of columns: first, location (IL - Israel, BR - Brazil, SA - Saudi Arabia, United States, New York City - US-NY); second, browser language (he - Hebrew, pt - Portuguese, ar - Arabic, en - English); third, browsing history (type of websites visited); fourth, number of bots deployed per type.}
  \Description{}
  \label{scheme_methodology}
\end{figure*}

\section{Methodology}
Our methodology relies on a Web crawling system, specifically developed to explore how search engine results may change for different user profiles. It enables the execution of Web crawlers (\textit{bots}) that can mimic typical online user behavior, be deployed to different locations, and make queries that are either related to the Israel-Palestine conflict or more general, non-sensitive queries. Figure \ref{scheme_methodology} summarizes our methodology for developing and implementing this querying and auditing system. It is divided into three main stages: \textit{1)} design of each bot, including its incremental feature design and deployment, \textit{2)} audit of search engine results, and \textit{3)} evaluation of search result differences in terms of URLs, content, and possible leaning.

\subsection{Bots and their Incremental Design}
The reasoning for using bots is twofold: (1) this approach allows precise control over key aspects of their online profiles. Unlike real user data, this systems makes it possible to isolate and analyze how specific factors, such as location or browsing history, influence search engine results. (2) Bots eliminate the ethical concerns of asking real users for their profiles or search histories, as this sometimes implies access to personal and sensitive information. Moreover, asking individuals to query search engines for potentially sensitive content, could affect their future online experience.

\subsubsection{Bots:} We developed each bot by extending the functionalities of OpenWPM\footnote{\url{https://github.com/openwpm/OpenWPM}} (Version 0.28.0), a community-trusted and flexible instrument for internet measurements and automating web browsing\cite{englehardt2016census}. 
Given the public policies of search engines (Table \ref{engines_policies}), previous results \cite{inconsistent_search_results, partisan_audience_bias, political_personalization}, and the geographical implications of the Israel-Palestine conflict, this work explores the influence of bots' (1) location, (2) browser language, and (3) browsing history on search engine results.
The bots were built using selenium and deploy version 123 of the Firefox browser, which can be parallelized and allows control of location and browsing settings such as language, cookies, and other forms of tracking, as detailed below:

\mypara{Location:} To change the location of the bots, Firefox settings were adjusted to use residential ISP proxies provided by Bright Data\footnote{\url{https://brightdata.com}}. %Similar to a typical VPN system, 
ISP proxies act as intermediate remote services for Internet access, masking the original source of the HTTP request. Using this system, the bots were placed in one of four locations: Israel (IL), Saudi Arabia (SA), Brazil (BR), and New York, United States (US-NY). Israel was selected due to its direct involvement in the conflict, with a majority Jewish and Hebrew-speaking population. Since Bright Data did not offer ISP proxies for Palestine, Lebanon, or Iran, Saudi Arabia was chosen for its majority Muslim and Arabic-speaking population. New York and Brazil were chosen as not-directly involved controls: New York due to its sizable Jewish community \cite{pew2015religious}, and the U.S.'s strong support for Israel, and Brazil as a more neutral location \cite{pew2013brazil}, with less support for Israel, and having a predominantly Christian population.

\mypara{Browser Language:} To simulate Internet access with browsers in the predominant language of the aforementioned locations, the Firefox language settings for each bot location were adjusted accordingly: Hebrew for Israel, Arabic for Saudi Arabia, Portuguese for Brazil, and English for the United States.

\mypara{Browsing History:} Online activity and browsing history (website visits, content interactions, etc.) can be tracked through cookies placed by first- and third-party sites, or by the specific configuration of the device and browser used to access the internet (fingerprinting)~\cite{cookiless_monster, stop_tracking_me_bro, webbrowsing_fingerprinting}. By instructing the bots to visit different online content (and saving the resulting tracking data), we aimed to create distinct browsing histories outside of the search engine platforms.

We created two types of browsing profiles by having the bots visit two types of URLs: (1) of local news that contained one of the keywords ``Palestine,'' ``Israel,'' ``Hamas,'' and ``Netanyahu'' (conflict-related, or specific), or (2) of local news that included the keywords ``movie,'' ``health,'' ``well-being,'' ``dinner recipe,'' and ``sports'' (non-conflict related, or general). The words were translated to the used languages (Hebrew, Arabic, Portuguese), to collect local news in the most spoken language of the country.
In both cases, the URLs were obtained through Media Cloud\footnote{\url{https://www.mediacloud.org/}}, a free and community supported tool that allows the retrieval of news articles from keywords, while filtering for location and time. The two lists of news URLs were obtained for each of the four selected locations, from October 7, 2023 to March 30, 2024.

%Extending predefined functions of the OpenWPM tool, 
The bots were instructed to visit sequentially 20 randomly selected websites, from one of the two lists (Figure~\ref{scheme_methodology}, 1A). During these visits, the bots collected and saved all tracking data (such as cookies, JavaScript fingerprinting files, and HTTP requests -- Figure~\ref{scheme_methodology}, 1B). This data was then stored in Firefox's profiles, which contained the bots' full browsing history and tracking information. 

\subsubsection{Incremental Design and Deployment:} 
To study the impact of these three features (location, browser language, and browsing history) on search engine results, we deployed three types of bots with an increasing number of features. \textit{Type 1} are bots specified to be solely associated to one of the four locations, \textit{Type 2} are bots that have both different location and browsing languages, and \textit{Type 3} are bots that diverge in their location, browser language, and browsing history (Figure~\ref{scheme_methodology}). Importantly, the bots of \textit{Type 3} were deployed simultaneously with bots having no browsing history (equivalent to a bot of \textit{Type 2}) as a temporal control. The table on the right side of Figure~\ref{scheme_methodology} presents a summary of the features used by each type of bot, as well as the number (last column) deployed by type.

\subsection{Search Engine Audit} 
In our study, three search engines were audited - DuckDuckGo, Google and Yahoo. Google was chosen for being the most used search engine worldwide; DuckDuckGo for its emphasis on privacy; and Yahoo for its association with a popular email service and extensive claims of extensive personalization.  

\subsubsection{Audit:} In each search engine audit, multiple bots -- each associated with different online user characteristics (see Table in Figure~\ref{scheme_methodology}) -- are simultaneously deployed on the same search engine. These bots input identical queries and collect the resulting URLs along with their rankings. This process is repeated across various queries and applied to the three search engines. To minimize the risk of carry-over effects (where previous searches could influence current results~\cite{personalization_web_search}), each audit begins with the same initial user profiles, launched in fresh browser instances.

Three audits were conducted between March 30 and May 4, with each audit lasting an average of two days and involving the deployment of bots of all types (Appendix, Figure~\ref{timeline}). Notably, in audits using Type 2 (location + browser language) and Type 3 (location + browser language + browsing history) bots, search engines automatically prioritized local results based on the language. 

To account for potential temporal fluctuations in ranking algorithms, two strategies were employed: (1) conducting simultaneous searches on the same search engine to minimize short-term variations, and (2) performing audits at least twice over distinct time points to address longer-term changes. 

\subsubsection{Search engine queries:}
Queries were split into two categories: ``general'' (denoted by the solid columns in the figures presenting our results, i.e., Figures~\ref{fig:main_plot} and~\ref{fig:rbo_per_category_website}) and ``conflict-specific'' (dashed columns in said figures). The first category comprises 27 popular and conflict-independent queries such as ``Home workout routines'', ``Popular books 2024'', ``how to tie a tie''. (Table~\ref{tab:queries_long_table}, column 2). 
The second category, specific to the conflict (e.g., ``military complex Al-Shifa hospital,'' ``Hamas rapes''), were defined in collaboration with the NGO HoneyComb, an investigative journalism association that tracks controversial online content, including disinformation. HoneyComb provided 10 distinct topics that were expanded using ChatGPT 3.5, to generate 27 specific queries (Table~\ref{tab:queries_long_table}, column 1).

A total of 27 queries per group (general and conflict-specific) were executed during the search engine audits for both Type 1 and Type 2. The audit deploying Type 3 bots performed a subset of 10 queries of the original set (Appendix, Table \ref{tab:queries_long_table}, text in bold). To maintain statistical integrity, each comparison of results ensured a consistent number of bots (accounting for occasional IP failures, Appendix \ref{number_successful_ports}) and queries per location (Appendix \ref{tab:number_queries_successful}).

\subsection{Quantifying Differences in Search Results}

After collecting search engine results and their respective rankings, the differences across different bot profiles were analyzed. This analysis was conducted in the three stages described next.
%: (1) Quantifying differences in URLs, (2) Quantifying differences in content, and (3) Measuring leaning of the search results.

\subsubsection{Quantifying Differences in URLs:}
To compare two lists of URLs returned by search engines queries (i.e., [website\_1.com, website\_2.com, ...] vs. [website\_3.com, website\_1.com, ...]), both the provided links and their order was considered, as users typically prioritize top results over lower-ranked ones \cite{Fay2024b}. The Rank-Biased Overlap (RBO) metric \cite{similar_measure} was used, which calculates how similar two lists are (0 totally different, 1 equal), considering the order of the items and different list sizes \cite{similar_measure}:

\[
\text{RBO}(S, T, p) = (1 - p) \sum_{d=1}^{\infty} p^{d-1} \cdot \frac{|S_{1:d} \cap T_{1:d}|}{d}
\]
where $S$ and $T$ are two ranked lists, $d$ is the rank and $p$ determines the top-weightiness of the metric. For our analysis, $p$ was set to 0.7, meaning that the first 10 results account for 99\% of the total value of the calculated RBO. The differences between two lists of search engine URLs results were calculated as $D = 1 - RBO$. Other metrics, such as edit distance and intersections of lists were also applied and gave similar results. 

\subsubsection{Quantifying Differences in Content}
To account for the possibility that the bots are receiving similar results through different local or language specific URLs (e.g.wikipedia.il vs. wikipedia.br), we classified all domains encountered (by the three types of bots, for both types of queries) in broad categories (e.g. Entertainment, News, etc.). The classification was initially done automatically by ChatGPT-4o, and then manually verified by one of the researchers for all domains (Appendix, Table \ref{tab:categories_category}).
%to classify all these domains according to their type and category . 
After this classification was performed, the same measurement, $D = 1 - RBO$ was computed, but considering lists of websites categories ([[News\_1, Political\_1, ...] vs. [Political\_1, News\_1...]), instead of the URL lists. 

\subsubsection{Quantifying Differences in Leaning}
To quantify whether the bots received results with particular leanings regarding the conflict, we focused on the URLs from domains classified as ``News'', for the 10 searches that were common to bots of all three types (see Appendix, Table \ref{tab:queries_long_table}). These articles were scraped, all domain-related information removed, and their title and content collected and automatically translated to English, to minimize possible coder bias from knowing the language or the news source.

Two approaches were used to quantify leaning on a common scale (pro-Israel, slightly pro-Israel, neutral, slightly pro-Palestine, pro-Palestine): asking the GPT-4o API (Appendix, Figure \ref{prompt_leaning}), and asking workers hired through the MTurk platform (Appendix, Figure \ref{mturker_instructions}). In both cases, the prompt was: ``Pro-Israel or pro-Palestine means that the text favors the narrative of that side of the conflict, or elicits sympathy towards it.'' MTurkers that classified texts had to pass two attention controls in a recruitment stage and during the entire classification task (per survey). For our analysis, we exclusively selected news texts that were classified by at least two MTurkers, with both attributing the same classification to the text.

% \subsection{Ethics}
% \input{ethical}

\section{Results}
%In this section we present the main results and findings of our study, divided by the steps of our incremental methodology and questions we aimed to address within each of the steps. 
We organized the results as answers to the main questions guiding our work. We ask if our bots (or proxy-users) see different results when querying search engines and, if so, how user's features affect those differences (section 4.1). This was done incrementally, by varying the level of bot customization: first varying only the location (\textit{Type 1}), then varying both location and browser language (\textit{Type 2}), and finally varying location, browser language, and browsing history (\textit{Type 3}). Next, we tested whether disparities in result URLs also correspond to meaningful differences in content (section 4.2) and/or content leaning (section 4.3) towards Israel or Palestine.

\begin{figure}[t]
    \centering
    \includegraphics[width=1\columnwidth]{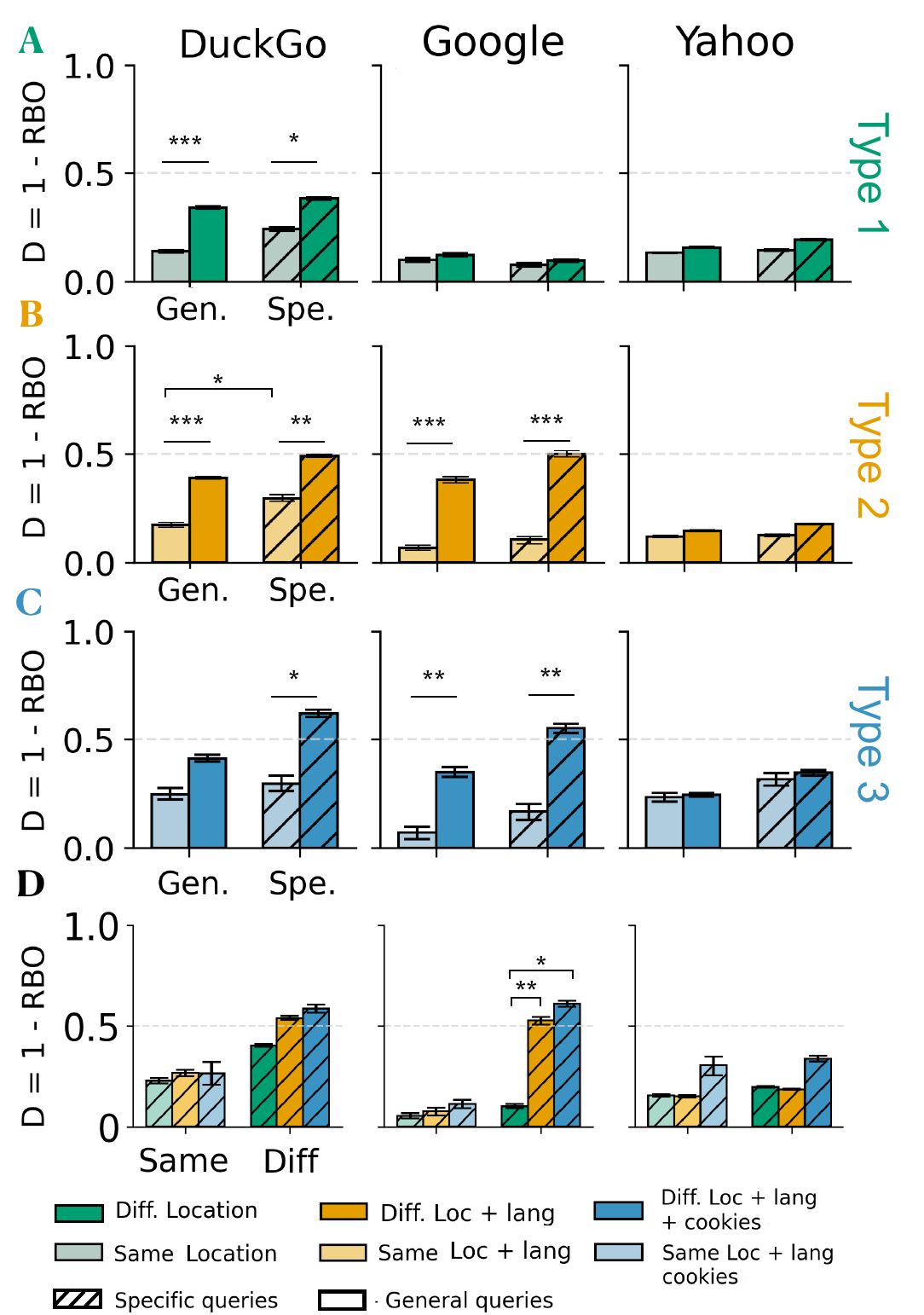}
    \caption{Panel A: Average difference ($D = 1 - RBO$) between search engine results for Type 1 bot pairs in the same location (light green) and different locations (dark green), for general (left) and specific (right, dashed) queries. Panel B: Average $D$ for Type 2 bots with the same (light yellow) and different (dark yellow) locations and browser languages, across query types and search engines. Panel C: Average $D$ for Type 3 bots with the same (light blue) and different (dark blue) location, browser language and browsing history. Panel D: Average $D$ for the 10 common specific queries across bot types, comparing bots with the same specifications (right) and different (left) across types (colors). Error bars represent 95\% confidence intervals based on the bootstrapped distribution. P-values were calculated with Mann-Whitney U test and adjusted with Bonferroni correction.*** denotes significant differences with p-value less than 0.001, ** less than 0.01, and * less than 0.05. Statistical details in Tables \ref{tab:p_values_adjusted_step_1}, \ref{tab:p_values_adjusted_step_2}, \ref{tab:p_values_adjusted_step_3}, \ref{tab:p_values_comparison_different_steps}.}
    \Description{}
    \label{fig:main_plot}
\end{figure}
%Do users see different resents for the same queries on search engines?  In this work we started by asking whether search engines results are impacted by location, browsing language and browsing history and whether this impact is more pronounced for queries of such a sensitive topic as the Israel-Palestine conflict or for general common queries. 
%In the next section, we detail our results. 

\subsection{Quantifying the Impact of Profiles and of Queries on Search Engine Results}
\subsubsection{Does geographical location influence Search Engine Results?}
\hfill\\
As discussed in the methods and summarized in Table \ref{engines_policies}, search engines typically report customizing results based on location or proxy location. This is useful when searching for a restaurant but might prove problematic when queries are on geographical conflicts. To examine how location affects search engine results, we deployed \textit{Type 1} bots in four distinct regions: Israel (IL), Saudi Arabia (SA), United States (US), and Brazil (BR) - details in the Methods. For each location, the bots queried the three search engines, and gathered the page results (URLs and their respective order) for every query, grouped as either conflict-specific or general.

Figure \ref{fig:main_plot}, panel A (green), summarizes these results, by comparing $D = 1 - RBO$ measurements (horizontally) for each search engine (vertically), with a high $D$ meaning that (the top) results were very different. Light green corresponds to the average $D$ values for bots deployed from the same location (e.g. one in BR bot compared to other BR bot) and dark green the same comparison but for bots deployed from different locations (e.g. BR bots vs US bots). 
%As the order in which the results are displayed is important (\cite{Fay2024b}), $D$ is weighed to prioritize the top results and a low  Then, the most common URL was identified and the second row, panel B, depicts the proportion of results that matched that most common result in each rank position. 
We found a tendency of larger variation between locations (dark green) than within locations (light green), the  difference being more pronounced for DuckDuckGo (p-values in Appendix, Table \ref{tab:p_values_adjusted_step_1}). These differences are even larger for queries related with the conflict, such as "Al-Shifa hospital" and "Hamas rapes" (dashed columns) when compared to more general searches, such as "Home workout routines" or "Popular books 2024" (solid columns). These findings persist even when controlling for query length (Appendix, Figure \ref{rbo_queries_same_size}), and applying other metrics, such as edit distance and intersection of lists (Appendix, Figure \ref{fig:other_metrics_step_1}). 
%Specifically, when comparing the values of $D=1 - RBO$ for "general" and "specific" queries with the same number of words, the same patterns hold. This rules out the possibility that the higher $D$ values observed for the "specific" group of queries are simply due to their longer average length.

%\newtcolorbox{mybox}[1]{colback=gray!5!white,
%colframe=lightgray!75!black,fonttitle=\bfseries,
%title={#1},
%boxsep=1mm, % Decreases the internal padding
%left=1mm, right=1mm, top=1mm, bottom=1mm % Sets specific padding values
%}

% \begin{mybox}{Finding 1}
% As anticipated, based on search engine customization policies, location significantly influences results. Importantly, this effect is more pronounced for conflict-related queries.
% \end{mybox}

\begin{takeaway}
\takeawaytitle{} As anticipated, based on search engine customization policies, location significantly influences results. Importantly, this effect is more pronounced for conflict-related queries.
\end{takeaway}

% \subsection{The impact of increasing bots profiling on search engine results}

% In this work, we aimed to test how different features of online profiles affect search engine results. We did this by gradually adding features to the bots' profiles when they made the searches. The goal was to make the bots more like real users and simultaneously measure how each new feature influenced the personalization of search engine results. As detailed in the Methods after testing the impact of location in search engine results (step 1) we added browser language (step 2) and browser history to our bots (step 3). The results are detailed in the following sections. 

\subsubsection{Does browser language impact search engine results?}
\hfill\\
In the second phase of our experiment, we configured \textit{Type 2} bots not only to be deployed from different locations, but also to interact with the search engines using the primary language of each chosen location (see Methods). These bots performed the same "general" and "specific" queries as before, under the hypothesis that this combination would further intensify the customization of search outcomes.

Figure \ref{fig:main_plot}, panel B (yellow), shows this: changing browser language leads to a greater divergence in search results across all search engines, particularly when comparing bots from different locations (light versus dark yellow columns) and, again, these variations are even more pronounced for conflict-related queries (dashed versus solid columns), with the data from Google revealing a particularly notable increase in the displayed URLs. Results for edit distance and list intersection can be seen in Figure \ref{fig:other_metrics_step_2}.

% \begin{mybox}{Finding 2} 
% The combination of location and browser language increases the personalization of search results, especially for Google. This effect is more prominent for conflict-related queries. 
% \end{mybox}

\begin{takeaway}
    \takeawaytitle{}
    The combination of location and browser language increases the customization of search results, especially for Google. This effect is more prominent for conflict-related queries.
\end{takeaway}

\subsubsection{Does browsing history impact search engine results?}
\hfill\\
It is debatable the extent to which search-engines personalize the displayed content based on the user's past browsing experience, and the studied search engines claim not to do it as long as users are not logged in (Background and Table \ref{engines_policies}). To test this, we directed the bots to visit (and collect cookies from) twenty websites, either related or unrelated to the conflict, prior to querying the search engines (\textit{Type 3} bots). If search engines do not personalize based on browsing history, these cookies should have no impact, and the results would mirror those of \textit{Type 2} (location and language) bots.
However, Figure \ref{fig:main_plot}, panel C (blue) shows that the average $D$ is even higher across all platforms, particularly for different locations (light versus dark blue) and for conflict-related queries (dashed versus solid columns). This held true even for DuckDuckGo, which claims not to use any user data, and for Google, despite the bots having no Google accounts. These findings remained consistent even when we compared the average $D$ values for only the 10 queries performed in this final experimental step (Figure \ref{fig:main_plot}, panel D, and Appendix, Figures \ref{fig:comparison_steps}).

Figure \ref{fig:main_plot}, panel D summarizes these findings: including browsing history (\textit{Type 3}, blue) increases variability in $D$, when compared to either \textit{Type 1} (green), or \textit{Type 2} (yellow) bots, and even more so for bots deployed in different locations (lighter versus darker colors). However, these differences were only significant in the case of Google (Appendix, Table  \ref{tab:p_values_comparison_different_steps} and \ref{table:stat_results}), possibly because of the very short, and not very nuanced, browsing profiles. Results for edit distance and list intersection can be seen in Figure \ref{fig:other_metrics_step_3}. 
%Future research should focus on creating more comprehensive and nuanced browsing history profiles. Nonetheless, it is noteworthy that, for Google, the comparisons of results of bots with different browsing histories were significant. 

% \begin{mybox}{Finding 3} Despite claims of limited or no personalization based on user information, browsing history influences the search engine displayed results, particularly for conflict-related queries. \end{mybox}

\begin{takeaway}
\takeawaytitle{}
Despite claims of limited or no personalization based on user information, browsing history influences the search engine displayed results, particularly for conflict-related queries.    
\end{takeaway}

% Once again we asked whether these differences in URLs also corresponded to differences in the topology of the results, that is, that search engines were not only retrieving different URLs possibly corresponding to more local sources, but whether they were actively retrieving different pages topologies. In Figure \ref{fig:rbo_per_category_website}, panel C, we observe that when we calculate the $D = 1-RBO$ for categories of websites and not URLs for Step 3, that differences become even more significant and once again, particularly, for the "specific" group of queries. 

\subsection{Differences in Search Engine Suggested Content}

One important question is whether the observed differences, especially between \textit{Type 1} and \textit{Type 2} bots, are mostly do to the search engines directing users to very similar content but in different local domains, which would appear as different URLs (for example, similar healthy recipes in different languages). To test this, we automatically classified all identified domains by category (e.g. News, E-Commerce Sports - see Methods and Appendix, Table \ref{tab:categories_category} and Figures \ref{categories_websites_general}, \ref{categories_websites_specific}) and recalculated $D$ for each category of websites. 

\begin{table}[t]
    \caption{Unique domains found by the bots in Step 1 of the experiment by search engine and for each category of queries.}
    \label{tab:categories_websites}
    \centering
    \begin{tabular}{p{1.2cm} p{1.2cm} p{4.3cm}} % Ensure width units (cm)
    \toprule
    Search\newline Engine & Query\newline Type &  Top 3 categories of \newline domains (\% prevalence) \\
    \midrule
    \multirow{2}{*}{\small DuckGo} & \small general & \small Lifestyle (26\%), Health (18\%), Entertainment (13\%)\\
    \cline{2-3}
    & \small specific & \small News (83\%), Reference (7\%), Education (3\%)\\
    \cline{1-3}
    \multirow{2}{*}{\small Google} & \small general & \small Lifestyle (16\%), Health (14\%), Business (10\%)\\
    \cline{2-3}
    & \small specific & \small News (61\%), Reference (17\%), Non-Profit (6\%)\\
    \cline{1-3}
    \multirow{2}{*}{\small Yahoo} & \small general & \small Lifestyle (26\%), Health (18\%), Entertainment (13\%)\\
    \cline{2-3}
    & \small specific & \small News (82\%), Reference (8\%), Education (4\%)\\
    \bottomrule
    \end{tabular}
\end{table}

\begin{figure}[t]
    \centering    \includegraphics[width=1\columnwidth]{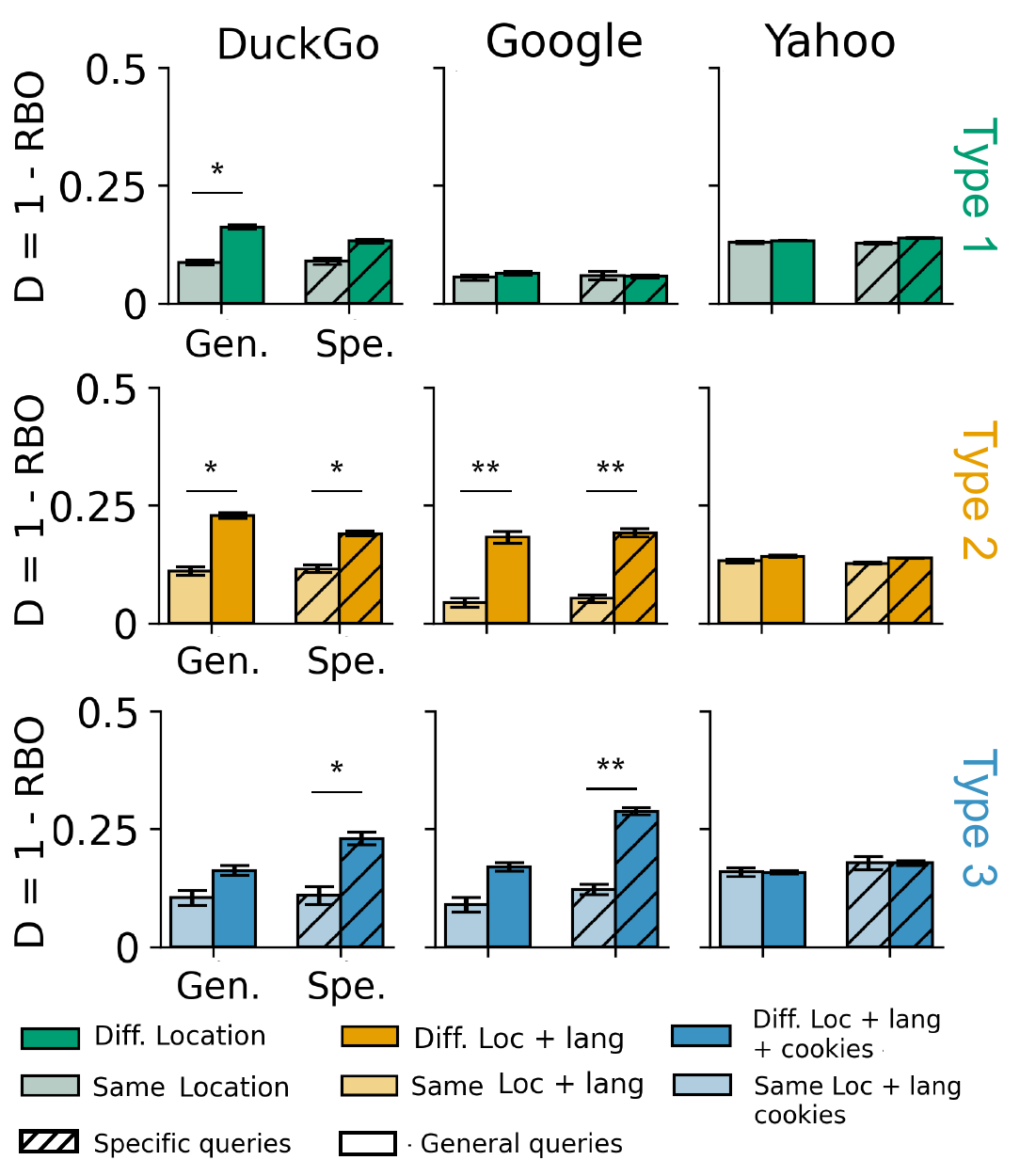}
    \Description{}
    \caption{Comparisons of ranked categories of websites for the three types of bots as in Figure 2. Only search results classified with a maximum of four different categories were considered, controlling for the number of queries per group. Error bars represent 95\% confidence intervals based on the bootstrapped distribution. P-values were calculated with Mann-Whitney U test and adjusted with Bonferroni correction.** denotes significant differences with a p-value less than 0.01 and * less than 0.05. Statistical details in Appendix, Tables \ref{tab:p_values_adjusted_categories_step_1}, \ref{tab:p_values_adjusted_categories_step_2}, \ref{tab:p_values_adjusted_categories_step_3}.}
    \label{fig:rbo_per_category_website}
\end{figure}

As the number of categories is much smaller than the initial number of URLs (or even domains), the variability measured by $D$ is expected to be much smaller. However, Figure \ref{fig:rbo_per_category_website} shows that there are still important differences, except in the case of Yahoo. For Google and DuckDuckGo, we find larger $D$ values between bot profiles than within them (light versus darker colors), particularly for those that have collected cookies (blue colors) and even more so in the case of conflict-related queries (dashed columns). The same is verified for edit distance and list intersection (Figures \ref{fig:other_metrics_step_2}, \ref{fig:other_categories_step_2}, \ref{fig:other_categories_step_3}). Interestingly, these differences are still higher for the "specific" category of websites even when a single category, "News", (Table \ref{tab:categories_websites} and Appendix, Figure \ref{categories_websites_specific}) dominates the results. Together, this indicates that searches from different bot profiles retrieve not only different content, but also that this content is sufficiently different to be classified into broad, distinct categories.

%the proportion of results per website category is significantly smaller for the "general" results compared to the "specific" results (see Table \ref{tab:categories_websites}). This outcome is expected, as the topics in the general query category are considerably more diverse than those in the specific category (refer to Appendix, Table 5). 

%Recalculating $D$ (as shown in Figure \ref{fig}) to focus solely on domain categories reduces variance, resulting in a smaller y-scale than in Figure \ref{fig:main_plot}. Importantly, more differences exist between bot profiles than within them, indicating that searches from different bot profiles retrieve varied content and exhibit distinct topologies. Furthermore, the larger variance associated with general queries remains evident, particularly among Type 3 bots (in blue), despite the reduced variance in categories for this group. These analyses demonstrate that bots often return different content for identical queries based on their apparent profile.

% \begin{mybox}{Finding 4}
% Distinct bot profiles are directed to URLs with different content. In the case of Google and DuckDuckGo, these differences are particularly pronounced for conflict-related, especially for bots that have collected cookies.
% \end{mybox}

\begin{takeaway}
    \takeawaytitle{}
    Distinct bot profiles are directed to URLs with different content. In the case of Google and DuckDuckGo, these differences are particularly pronounced for conflict-related, especially for bots that have collected cookies.
\end{takeaway}

% \subsection{The impact of increasing bots profiling on search engine results}

% In this work, we aimed to test how different features of online profiles affect search engine results. We did this by gradually adding features to the bots' profiles when they made the searches. The goal was to make the bots more like real users and simultaneously measure how each new feature influenced the personalization of search engine results. As detailed in the Methods after testing the impact of location in search engine results (step 1) we added browser language (step 2) and browser history to our bots (step 3). The results are detailed in the following sections. 

% We then asked if this variation in displayed URLs also corresponded to differences and content. The results displayed in \ref{fig: rbo_per_category_website}, panel B, show that our observations hold even when we calculate $D = 1-RBO$ for categories of websites. That is, search engines are presenting more divergent content to bots located at different countries and searching with different browser languages, with this content diverging more for searches specific to the conflict. 

\subsection{Leaning of Search Engine Results}

\begin{figure}[t]
    \centering
    \includegraphics[width=1\columnwidth]{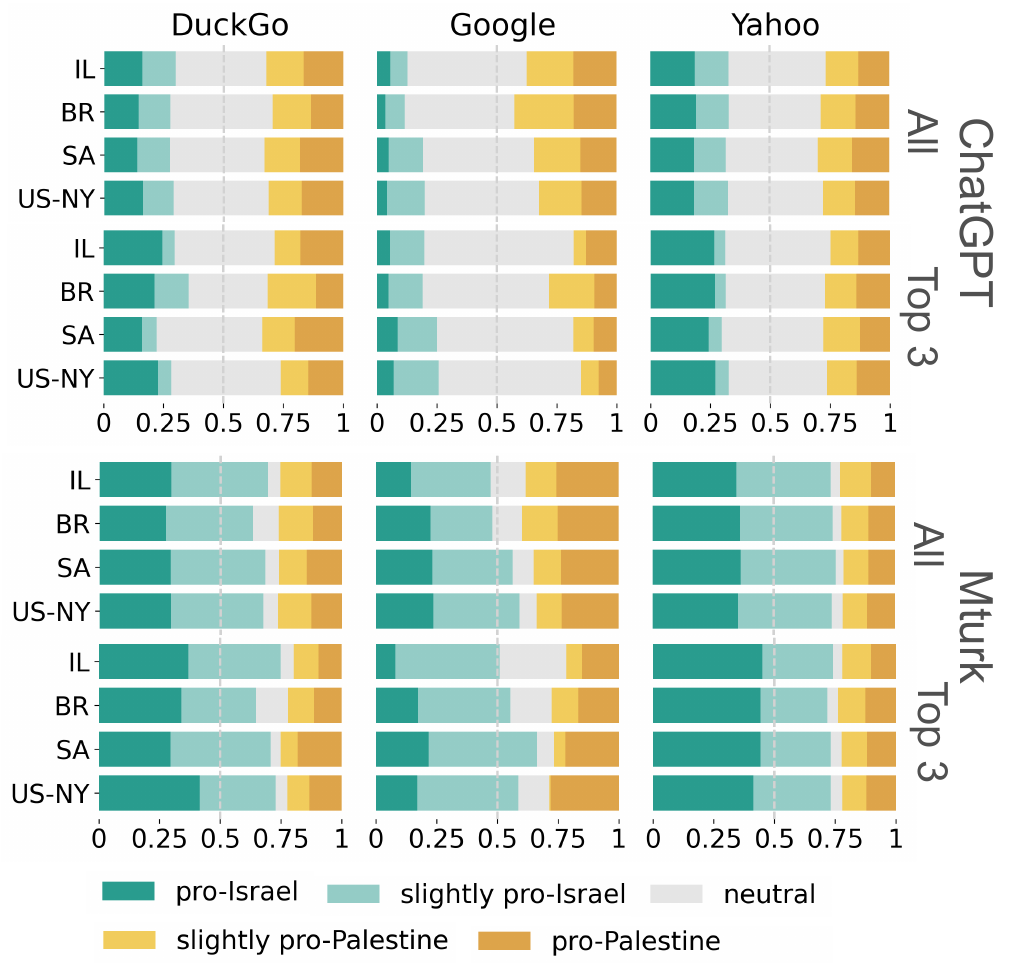}
    \caption{Proportion of news articles classified as being "neutral" (grey), "pro-Israel" and "slightly pro-Israel" (blue tones) and as being "pro-Palestine" and "slightly pro-Palestine" (yellow tones), for each country (y-axis) and search engine. The top panel shows ChatGPT classification considering all news results and the news present in the first 3 results of the pages. Bottom panel, 
    same, but with human classification.}
    \label{leaning}
    \Description{}
\end{figure}

Even if the results and the page contents are different (and even if these differences increase when the profile of the bots becomes more detailed), it could still be argued that there is so much randomness in the algorithmic decisions of search-engines, that the observed variance would probably be inconsequential. Conversely, if the algorithms of search engines use information such as location and past browsing experience to customize political content, this could further amplify preexisting views. % and, as been extensively argued in the case of social networks, lead to the creation of "Echo-chamber" effects (REF). 
Therefore, we asked whether the differences in content, particularly in the case of the conflict-specific URLs, also reflect differences in leaning. We focused on the "News" category of websites, as it corresponds to the majority of results in the conflict-specific group of queries (Table \ref{tab:categories_websites}), and evaluated their leaning using both human coders and ChatGPT (Methods and Appendix, Figure \ref{prompt_leaning} and \ref{mturker_instructions}). 
As Figure \ref{leaning} shows, ChatGPT (top) classifies as neutral many more news than the human coders (bottom), with the MTurkers classifying more content as pro-Israel than pro-Palestine (more analysis on the level of agreement between chatGPT and MTurkers in Appendix, Figure \ref{heat_map}). Focusing on ChatGPT results, interestingly, we observe that as generally more news are classified as pro-Palestine (yellow - top panel), for all search engines, when we analyze the leaning of news present in the top 3 results, the number of pro-Israel significantly increases (particularly for Google). With Google having a tendency of showing even less pro-Palestine results for IL and US in the top 3 results. The same happens for DuckDuckGo. Impressively, Yahoo is quite consistent about the number of leaning articles showed to bots with different locations. 

When we focus on MTurk classifications (bottom panel), we observe the same tendency of decreasing the number of pro-Palestine articles in the top 3 results, with this behavior being more pronounced for IL for both DuckduckGo and Google. 

%\begin{mybox}{Finding 5}
%Bots located at different locations receive content with different %leanings regarding the conflict, according to both ChatGPT and %Mturkers. Duckduckgo and Google tend to show more pro-Israel content in their top 3 results, with a decrease in pro-Palestine, and this decrease is dependent of the bot location. 
%\end{mybox}

\begin{takeaway}
\takeawaytitle{}
Bots at different locations receive content with different leanings regarding the conflict, according to both ChatGPT and MTurkers. Duckduckgo and Google tend to show more pro-Israel content in their top 3 results than in All, with a decrease in pro-Palestine, dependent on the bot location.
\end{takeaway}

\section{Discussion}
Here, we discuss some of the weaknesses and threats to validity: %Some of the strengths of this study, such as not using human data and using an incremental, combinatorial approach, also work as weaknesses as detailed below:

\mypara{1. Non-human queries}: Despite our best efforts, the bots do not perfectly mimic humans and, if identified as bots, the search engines might handle the their queries differently than they would the human-generated traffic. To minimize this, we coded the bots to act more human-like (introducing delays, natural ``typing'', avoiding VPN farms, etc.) and we have reason to believe that they were only rarely identified as such (e.g. we received few ``captcha'' requests and only some queries failed). Also, the bots have none (\textit{Type 1} and \textit{Type 2}) to limited (\textit{Type 3}) browsing history, and this could limit personalization of the results.  However, that even such a conservative approach managed to reveal significant differences for conflict-specific queries, should raise important alarms, as these are likely to be even higher for human queries. In the future, audits could enhance the realism of the bots by increasing browsing history (breadth and size) and adding features such as search language or social media profiles.
    
\mypara{2. Combinatorial complexity}: The incremental and combinatorial system design led to a rapid increase in the number of variables, making exhaustive testing complex and expensive. \textit{Type 3} bots only queried approximately one third of the original list and we used a conservative approach to their selection: the selected 10 queries were the ones for which \textit{Type 1} and \textit{Type 2} bots had the smallest differences in $D$, between locations. Thus, it is possible that if other queries had been chosen the observed effects would have been larger, but it is unlikely they would have been smaller.

\mypara{3. Geographical IP limitations}: No IP addresses were used from countries historically more supportive of Palestine, such as Lebanon and Iran. Our choice of Saudia Arabia was not ideal, particularly as most searches were done in English, and this omission may limit the identification of clearly pro-Palestine content in this region. Even though we used residential proxies, typically very hard to identify, occasionally they failed for unknown reasons (see Appendix \ref{number_successful_ports}). 
    
\mypara{4. Dynamic nature of search engines}: That the algorithms of search engines are both fast-evolving and proprietary means that even minor variations in the timing of our audits could yield different results. To control for such temporal fluctuations, particularly during such a fast-evolving situation, we repeated the same queries for stateless Type 2 bots three times and observed no temporal trend (Appendix, Figures \ref{time_control_1}, \ref{time_control_2}, \ref{time_control_3}, \ref{time_control_4}). Future studies should update testing protocols more frequently and include more queries, search engines, and LLM-based interfaces.
 
\mypara{5. Analysis of leaning}: There is no objective metric of leaning and there were significant differences in classification between ChatGPT and human coders. In particular, ChatGPT was more likely to consider as neutral, news that the humans identified as being more favorable to Israel. Future work should have more coders, from different locations and/or include expert coders.

\section{Conclusions and Future Work}
This paper presents a novel methodology for auditing search engine results using web crawlers designed to mimic different ``human-like'' online profiles. By focusing on the Israel-Palestine conflict and examining how factors such as location, browser language, and browsing history affect search results across Google, DuckDuckGo, and Yahoo, we demonstrated that there are significant differences in search engine results, as these features vary.
%personalization plays a significant role in shaping the information landscape for sensitive or contentious topics.

Our findings confirm that location is a critical factor in influencing search engine results, as has been shown in previous studies \cite{inconsistent_search_results, personalization_web_search}. Importantly, tailoring search results to the user's location is typically considered ``neutral" or not worrisome, but that is clearly not the case for geo-political conflicts: that individuals across political borders are shown different content (possibly with even different political leaning) could have serious consequences, difficult to anticipate and measure.

We also found that combining location with other factors like browser language and browsing history increases the variability in results, particularly for conflict-related queries. Importantly, this effect is not limited to differences in URLs but extends to the types of content displayed, such as news or political content. Contrary to some earlier studies, our results revealed that browsing history does induce notable differences, particularly in the case of Google, and this may further amplify the effects of information bubbles or echo chambers. Moreover, we observed that search engines may not fully adhere to their stated personalization policies. Google and DuckDuckGo, despite their claims not to personalize results based on "naive users" browsing history, displayed significant differences in the results, whereas Yahoo showed the least variation, despite claiming the opposite policy. Future work could explore the nuances of how browsing history impacts customization, particularly by disentangling how and which cookies lead to particular results.

Currently, the way forward in the online tracking space is uncertain, with Google leading a shift away from third-party cookies and then seemingly turning around \cite{google_privacy_sandbox}. It is important that regardless of the technologies employed,  audits such as this remain possible. In the European Union, the Digital Services Act specifically states that very large online search engines must provide access to data and/or code deemed necessary to monitor and guarantee compliance on areas such as privacy protection and personalization. This regulation also includes a mechanism for researcher access and academic auditing around systemic threats. Using such a mechanism, it should be possible to extend this methodology or to request access to relevant information on algorithmic recommendations. Focusing on other controversial topics could help further investigate the role of search engines in shaping public discourse across key political or cultural issues. Additionally, the development of more sophisticated tools for measuring the political leanings of search results could provide deeper insights into the biases search engines may introduce, particularly in global conflicts.

\begin{acks}
We thank members of the Social Physics and Complexity (SPAC) group at LIP for comments and critical reading of the manuscript. We thank HoneyComb for support in identifying meaningful topics for this study and BrightData for generously providing proxy services, free of charge. This research was partially funded by ERC Stg FARE (853566) and ERC PoC FARE\_Audit (101100653), both to JGS, and by FCT PhD fellowship (2022.12547.BD) to ID.
\end{acks}
\balance
\bibliographystyle{ACM-Reference-Format}
\bibliography{refs}

\clearpage

\appendix
\supplementarysection
\section{Supplementary Description of the Data collected}

\subsection{Data Collection Timeline}
\begin{figure}[ht]
    \centering \includegraphics[width=1\columnwidth]{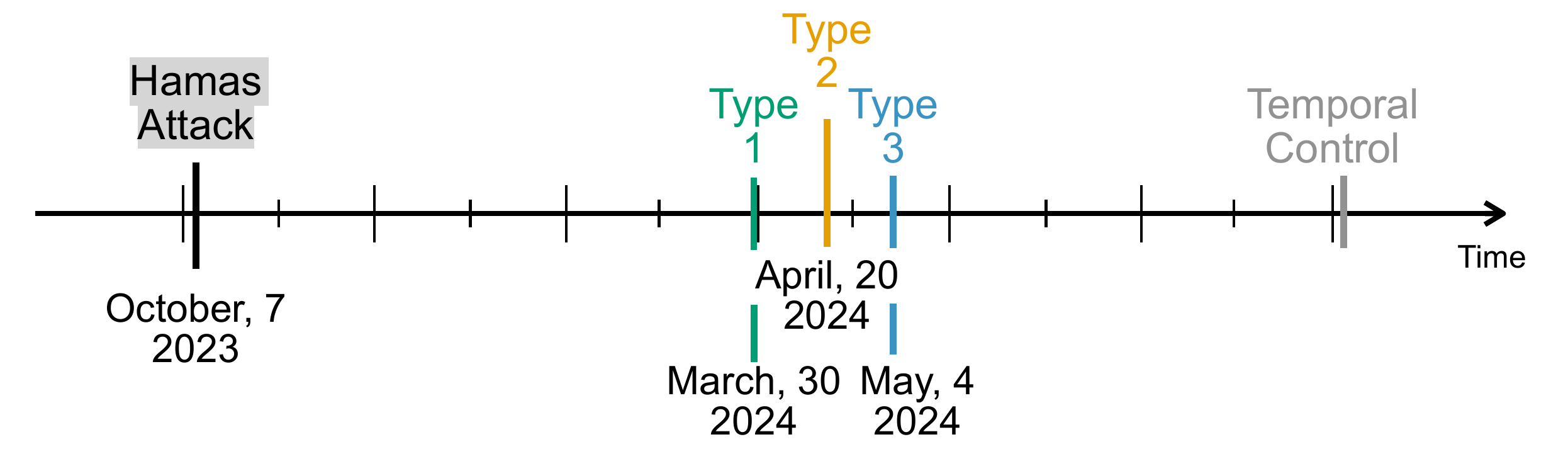}
    \caption{Temporal alignment of deployment of Type 1, Type 2 and Type 3 bots, and the temporal control.}
    \Description{}
    \label{timeline}
\end{figure}

In Figure \ref{timeline}, we show the moments when data collection began for each Type of webcrawlers used in the experiment.

\subsection{Queries used in the experiment}

Table \ref{tab:queries_long_table} shows all queries used in this experiment. Notice that Type 3 bots performed a smaller number (10) of the original number of queries (gray background). 

\begin{table}[h]
    \centering
    \caption{Queries used in the experimental setup. All specific and general queries shown here were used in deployment of Type 1 bots (Location only) and Type 2 (Location + Browser Language). Type 3 (Location + browser language + browsing history) only did queries in bold (10 per category).}
    \label{tab:queries_long_table}
    \begin{tabular}{|| p{3.7cm} | p{3.7cm} ||}
    \hline
    \textbf{Specific Queries} & \textbf{General Queries} \\
    \hline\hline
    \small military complex Al-Shifa hospital & \small How to tie a tie \\ 
    \hline
    \small Israel banned Olympics & \small Popular books 2024 \\
    \hline
    \small Tiktok antisemitism & \small \textbf{Home workout routines} \\
    \hline
    \small Israeli babies beheaded & \small best movies ever \\
    \hline
    \small \textbf{Hamas rapes} & \small How to grow indoor plants \\
    \hline
    \small strike Al-Ahli hospital & \small Uncommon hobbies to try \\
    \hline
    \small \textbf{Gaza tunnels} & \small Rare and endangered plant species \\
    \hline
    \small Palestine banned Olympics & \small \textbf{Experimental music genres}\\
    \hline
    \small Saint Porphyrios Orthodox Church Gaza & \small organic gardening tips \\
    \hline
    \small \textbf{Hamas} & \small how to be smarter \\
    \hline
    \small Supernova festival attack &  \small\textbf{ Financial planning tips} \\
    \hline
    \small Israel destroyed an orthodox church in Gaza & 
    \small \textbf{Benefits of meditation for mental health} \\
    \hline
    \small \textbf{Erdogan threatened to intervene and support Palestinians?} & \small Nutritional value of quinoa \\
    \hline
    \small Ukraine provided weapons to Hamas & \small How to improve sleep quality naturally? \\
    \hline
    \small Yemen has declared war against Israel & \small Interesting facts about dolphins \\
    \hline
    \small \textbf{Gaza} & \small \textbf{Guide to composting at home} \\
    \hline
    \small Palestinian nurse claims Hamas steals food and medicine from al-Shifa Hospital & \small Popular science fiction books 2024 \\
    \hline
    \small \textbf{Israel} & \small How to set up a home office? \\
    \hline
    \small \textbf{American troops have landed in Israel to help Netanyahu's war efforts} & \small \textbf{Effects of climate change on wildlife} \\
    \hline
    \small Orthodox church Gaza destroyed by Israel & \small Tips for growing herbs indoors \\
    \hline
    \small \textbf{Houthis} & \small underrated movies you must see \\
    \hline
    \small \textbf{Erdogan threatens to support Palestinians} & \small \textbf{Current trends in sustainable fashion} \\
    \hline
    \small Ukraine provides weapons to Hamas &  \small \textbf{How do electric cars work?}\\
    \hline
    \small Yemen declares war on Israel & \small \textbf{Healthy lunch ideas for work} \\
    \hline
    \small Palestinian nurse alleges Hamas theft from al-Shifa Hospital & \small How to invest in stocks for beginners? \\
    \hline
    \small \textbf{American troops land in Israel to aid Netanyahu's war} & \small \textbf{DIY home decor on a budget} \\
    \hline
    \small Middle East conflict & \small What is virtual reality and how does it work? \\
    \hline
    \end{tabular}
\end{table}

\subsection{Data Description}
In the deployment of each bot's type the first page of results was collected for each combination of search engine, bot, and query. However, not every audit (i.e., the collection of results for a query) was successful. Failures were often due to IP issues or problems with the xpaths of certain search engine results. In the analysis, only successful audits were considered. A successful audit was defined as the collection of at least four URLs from at least three different IPs within the same location for the same search engine and query. Since the failure of some audits introduced imbalances in the search engine results database, we ensured that each analysis used the same number of specific and general queries per search engine, as well as the same number of ports per location.  

Table \ref{tab:number_queries_successful}  presents the number of successful queries in each category for every search engine and location pair considered in the analysis. It is important to note that while the number of successful queries per search engine may vary for the same experimental step, we ensured that the number of specific and general queries was consistent within each search engine analysis.

\begin{table}[hb]
    \centering
    \caption{Number of successful queries within each query category per search engine.}
    \label{tab:number_queries_successful}
    \begin{tabular}{|p{2cm}|p{2cm}|p{2.5cm}|}
    \hline
    \textbf{} & \textbf{Search Engine} & \textbf{Number Queries}\\
    \hline
    \multirow{4}{*}{Type 1} & DuckDuckGo & 26 \\
    \cline{2-3}
    {} & Google & 27 \\
    \cline{2-3}
    {} & Yahoo & 27 \\
    \hline
    \multirow{4}{*}{Type 2} & DuckDuckGo & 27 \\
    \cline{2-3}
    {} & Google & 19 \\
    \cline{2-3}
    {} & Yahoo & 27 \\
    \hline
    \multirow{4}{*}{Type 3} & DuckDuckGo & 10 \\
    \cline{2-3}
    {} & Google & 10 \\
    \cline{2-3}
    {} & Yahoo & 10 \\
    \hline
    \end{tabular}
\end{table}

\begin{figure}[hb]
\includegraphics[width=1\columnwidth]{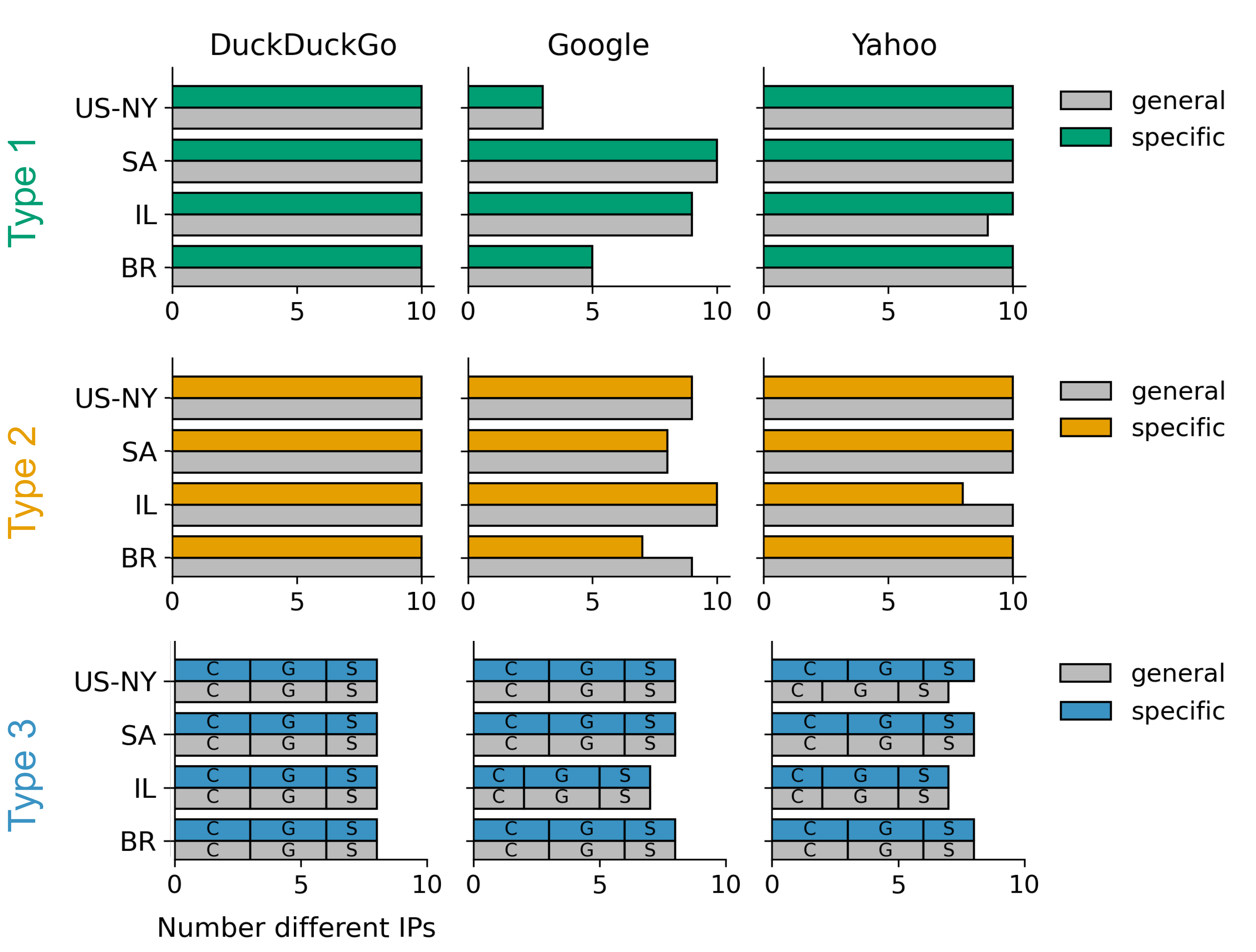}
    \caption{Number of successful ports per location across search engines and bot type. The gray bars represent the number of IPs successful in performing general queries, while the colored bars indicate those successful in specific queries. For Type 3 bots, each bar is further divided into the number of IPs trained with visits to conflict news websites (C), general news websites (G), and stateless bots (S).}
    \Description{}
    \label{number_successful_ports}
\end{figure}

Figure \ref{number_successful_ports} shows the number of successful IPs for each location and search engine. 10 ports of Type 1 and Type 2 were deployed initially. However, of Type 3, only 8 different IPs were used per location. Of these 8 IPs, 3 were associated with visits to news articles about the studied conflict (labeled as "C" in Figure \ref{number_successful_ports}, Type 3), 3 had a profile of regular news browsing history (labeled as "G"), and 2 were stateless bots ("S").

Table \ref{tab:number_results} contains a small description of the total number of URLs collected in each step of the experiment. 

\begin{figure}[H]
    \centering
    \includegraphics[width=0.8\columnwidth]{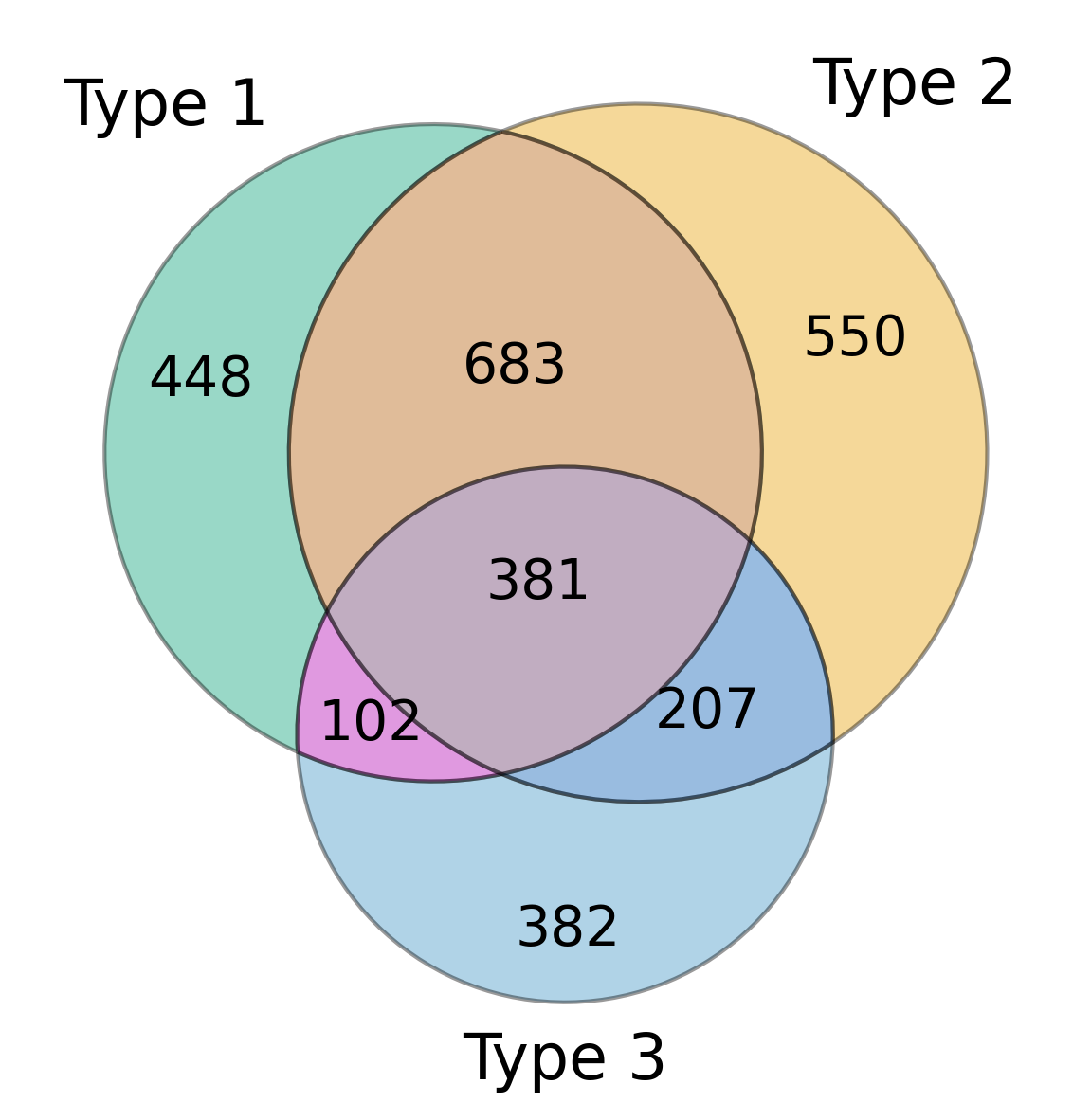}
    \caption{Venn diagram of unique URLs present in Type 1 - Location only (green), Type 2 - Location + browser languages (yellow) and Type 3 - Location + browser languages + browsing history (blue) and its intersections.}
    \Description{}
    \label{venn_unique_urls}
\end{figure}

\begin{figure}[H]
    \centering
    \includegraphics[width=0.8\columnwidth]{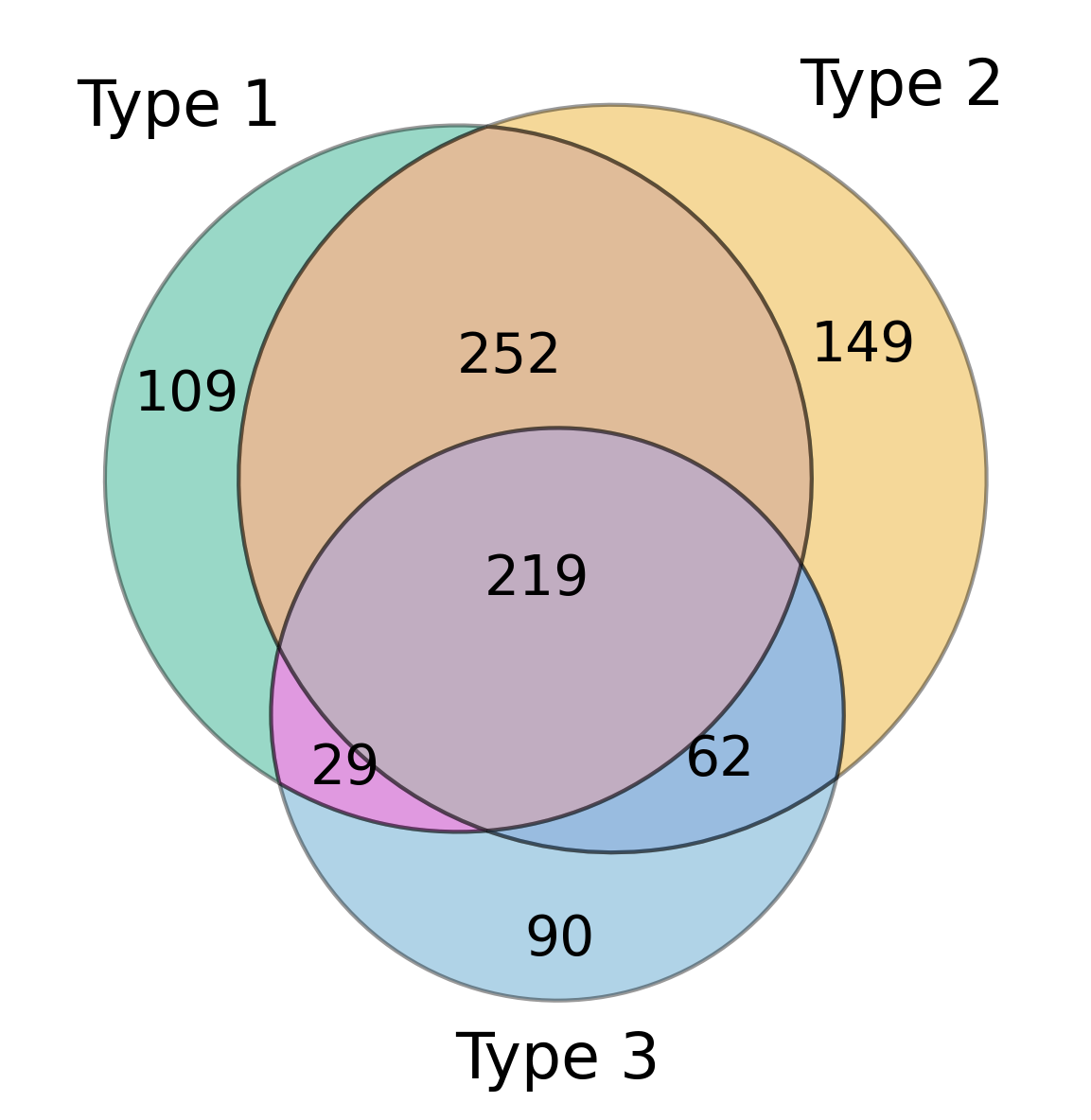}
    \caption{Venn diagram of unique domains present in Type 1 - Location only (green),  Type 2 - Location + browser languages (yellow) and Type 3 - Location + browser languages + browsing history (blue) and its intersections.}
    \Description{}
    \label{venn_unique_domains}
\end{figure}

\newpage
Some results were present to all bots Types. In the Venn diagram of Figure \ref{venn_unique_urls} we show the intersection of raw URLs across the types of bots.  In the Veen diagram of Figure \ref{venn_unique_domains} the intersection of websites domains across bot types.

\begin{table}[t]
    \centering
    \caption{Description of results (URLs) data base considered for analysis. N results corresponds to the number of total results, N unique results to the number of different URLs collected for all queries and N unique domains to the total number of different websites domains.}
    \label{tab:number_results}
    \begin{tabular}{|p{0.8cm}|p{1cm}|p{1.5cm}|p{1.5cm}|p{1.5cm}|}
    \hline
    \textbf{} & \textbf{Search \newline Engine} & \textbf{Query \newline category} & \textbf{N unique results} & \textbf {N unique domains}\\
    \hline
    \multirow{4}{*}{Type 1} & \multirow{2}{*}{DuckGo} & General & 547 & 335\\
    \cline{3-5}
    {} & {} & Specific & 597 & 104 \\
    \cline{2-5}
    {} & \multirow{2}{*}{Google} & General & 328 & 224\\
    \cline{3-5}
    {} & {} & Specific & 268 & 98\\
    \cline{2-5}
    {} & \multirow{2}{*}{Yahoo} & General & 244 & 168 \\
    \cline{3-5}
    {} & {} & Specific & 245 & 57\\
    \hline
    \multirow{4}{*}{Type 2} & \multirow{2}{*}{DuckGo} & General & 621 & 380\\
    \cline{3-5}
    {} & {} & Specific & 746 & 125 \\
    \cline{2-5}
    {} & \multirow{2}{*}{Google} & General & 304 & 216\\
    \cline{3-5}
    {} & {} & Specific & 338 & 119\\
    \cline{2-5}
    {} & \multirow{2}{*}{Yahoo} & General & 207 & 154 \\
    \cline{3-5}
    {} & {} & Specific & 208 & 51\\
    \hline
    \multirow{4}{*}{Type 3} & \multirow{2}{*}{DuckGo} & General & 226 & 158\\
    \cline{3-5}
    {} & {} & Specific & 379 & 66 \\
    \cline{2-5}
    {} & \multirow{2}{*}{Google} & General & 216 & 164\\
    \cline{3-5}
    {} & {} & Specific & 339 & 111\\
    \cline{2-5}
    {} & \multirow{2}{*}{Yahoo} & General & 108 & 83 \\
    \cline{3-5}
    {} & {} & Specific & 152 & 41\\
    \hline
    \end{tabular}
\end{table}

\newpage
\section{Additional Results and Analysis}

\subsection{Statistical Results}
The statistical analysis values present in \ref{fig:main_plot} are described in the the following tables \ref{tab:p_values_adjusted_step_1}, \ref{tab:p_values_adjusted_step_2}, \ref{tab:p_values_adjusted_categories_step_3}, \ref{tab:p_values_comparison_different_steps} in more detail. 

\begin{table}[ht]
    \centering
    \caption{Pairwise comparisons of RBO per query type and location for Type 1 bots across search engines, showing the respective p-values and Bonferroni-adjusted p-values. The Mann-Whitney U test was used for the comparisons, and significant adjusted p-values ($<0.05$) are highlighted in bold. The adjustment accounts for 12 comparisons across the three search engines.}
    \Description{}
    \label{tab:p_values_adjusted_step_1}
    \begin{tabular}{|p{1cm}|p{3cm}|p{1cm}|p{1.5cm}|}
    \hline
    \textbf{Search\newline Engine} & \textbf{Groups \newline compared} & \textbf{p-value} & \textbf{Adjusted \newline p-value} \\
    \hline\hline
    \multirow{4}{*}{DuckGo} & Same Location vs. Diff Location - General Queries & \textbf{$<<0.001$} & \textbf{$<<0.001$} \\
    \cline{2-4}
    & Same location vs. Diff location - Specific Queries & \textbf{0.002} & \textbf{0.02} \\
    \cline{2-4}
    & General vs. Specific - Same Location & \textbf{0.05} & 0.56\\
    \cline{2-4}
    & General vs. Specific - Diff Location & 0.88 & 10.52\\
    \hline \hline
    \multirow{4}{*}{Google} & Same Location vs. Diff Location - General Queries & 0.33 & 3.98 \\
    \cline{2-4}
    & Same location vs. Diff location - Specific Queries & 0.14 & 1.69 \\
    \cline{2-4}
    & General vs. Specific - Same Location & 0.21 & 2.48 \\
    \cline{2-4}
    & General vs. Specific - Diff Location & 0.25 & 2.96\\
    \hline \hline
    \multirow{4}{*}{Yahoo} & Same Location vs. Diff Location - General Queries & 0.07 & 0.86 \\
    \cline{2-4}
    & Same location vs. Diff location - Specific Queries & 0.02 & 0.29 \\
    \cline{2-4}
    & General vs. Specific - Same Location & 0.20 & 2.40\\
    \cline{2-4}
    & General vs. Specific - Diff Location & 0.11 & 1.41\\
    \hline
    \end{tabular}
\end{table}
\vspace{1cm}
\begin{table}[hb]
    \centering
    \caption{Pairwise comparisons of RBO per query type and location for Type 2 bots across search engines, showing the respective p-values and Bonferroni-adjusted p-values. The Mann-Whitney U test was used for the comparisons, and significant adjusted p-values ($<0.05$) are highlighted in bold. The adjustment accounts for 12 comparisons across the three search engines.}
    \Description{}
    \label{tab:p_values_adjusted_step_2}
    \begin{tabular}{|p{1cm}|p{3cm}|p{1cm}|p{1.5cm}|}
    \hline
    \textbf{Search\newline Engine} & \textbf{Groups \newline compared} & \textbf{p-value} & \textbf{Adjusted \newline p-value} \\
    \hline\hline
    \multirow{4}{*}{DuckGo} & Same Location vs. Diff Location - General Queries & \textbf{$<<0.001$} & \textbf{$<<0.001$} \\
    \cline{2-4}
    & Same location vs. Diff location - Specific Queries & \textbf{<<0.001} & \textbf{<<0.001}\\
    \cline{2-4}
    & General vs. Specific - Same Location & \textbf{<<0.001} & \textbf{<<0.001}\\
    \cline{2-4}
    & General vs. Specific - Diff Location & \textbf{0.04} & 0.47\\
    \hline \hline
    \multirow{4}{*}{Google} & Same Location vs. Diff Location - General Queries & \textbf{$<<0.001$} & \textbf{$<<0.001$} \\
    \cline{2-4}
    & Same location vs. Diff location - Specific Queries & \textbf{<<0.001} & \textbf{<<0.001} \\
    \cline{2-4}
    & General vs. Specific - Same Location & 0.10 & 1.30 \\
    \cline{2-4}
    & General vs. Specific - Diff Location & 0.10 & 1.15\\
    \hline \hline
    \multirow{4}{*}{Yahoo} & Same Location vs. Diff Location - General Queries & 0.27 & 3.30 \\
    \cline{2-4}
    & Same location vs. Diff location - Specific Queries & \textbf{0.02} & 0.25 \\
    \cline{2-4}
    & General vs. Specific - Same Location & 0.60 & 7.17\\
    \cline{2-4}
    & General vs. Specific - Diff Location & 0.12 & 1.48\\
    \hline
    \end{tabular}
\end{table}
\clearpage
\begin{table*}[t]
    \centering
    \caption{Pairwise comparisons of RBO per query type and location for Type 3 bots across search engines, showing the respective p-values and Bonferroni-adjusted p-values. The Mann-Whitney U test was used for the comparisons, and significant adjusted p-values ($<0.05$) are highlighted in bold. The adjustment accounts for 12 comparisons across the three search engines.}
    \Description{}
    \label{tab:p_values_adjusted_step_3}
    \begin{tabular}{|p{2cm}|p{7cm}|p{1.5cm}|p{2.5cm}|}
    \hline
    \textbf{Search Engine} & \textbf{Groups compared} & \textbf{p-value} & \textbf{Adjusted p-value} \\
    \hline\hline
    \multirow{4}{*}{DuckGo} & Same Location vs. Diff Location - General Queries & \textbf{0.03} & 0.37 \\
    \cline{2-4}
    & Same location vs. Diff location - Specific Queries & \textbf{0.001} & \textbf{0.01}\\
    \cline{2-4}
    & General vs. Specific - Same Location & 0.34 & 4.14\\
    \cline{2-4}
    & General vs. Specific - Diff Location & \textbf{0.01} & 0.14 \\
    \hline \hline
    \multirow{4}{*}{Google} & Same Location vs. Diff Location - General Queries & \textbf{$<<0.001$} & \textbf{0.009} \\
    \cline{2-4}
    & Same location vs. Diff location - Specific Queries & \textbf{$<<0.001$} & \textbf{<<0.001} \\
    \cline{2-4}
    & General vs. Specific - Same Location & \textbf{0.02} & 0.33 \\
    \cline{2-4}
    & General vs. Specific - Diff Location & \textbf{0.02} & 0.25\\
    \hline \hline
    \multirow{4}{*}{Yahoo} & Same Location vs. Diff Location - General Queries & 0.57 & 6.85 \\
    \cline{2-4}
    & Same location vs. Diff location - Specific Queries & 0.52 & 6.24 \\
    \cline{2-4}
    & General vs. Specific - Same Location & 0.14 & 1.69\\
    \cline{2-4}
    & General vs. Specific - Diff Location & 0.14 & 1.69\\
    \hline
    \end{tabular}
\end{table*}
\clearpage
\begin{table*}[b]
    \centering
    \caption{Pairwise Comparison of results from different bots, across search engines and query classifications with p-values and respective Adjusted p-values. Adjusted p-values are calculated using the Bonferroni method and when significant ($<0.05$) are highlighted in bold. The adjustment accounts for 36 comparisons counting with number of tests (6), per search engine and query groups.}
    \label{tab:p_values_comparison_different_steps}
    \begin{tabular}{|p{3cm}|p{2cm}|p{5cm}|p{1.5cm}|p{3cm}|}
    \hline
    \textbf{Search Engine} & \textbf{Query Type} & \textbf{Groups compared} & \textbf{p-value} & \textbf{Adjusted p-value} \\
    \hline
    \multirow{6}{*}{DuckDuckGo} & \multirow{6}{*}{General} & Type 1 vs. Type 2 - Same Location & 0.97 & 34.91 \\
    & & Type 1 vs. Type 3 - Same Location & 0.27 & 9.83 \\
    & & Type 2 vs. Type 3 - Same Location & 0.38 & 13.85 \\
    & & Type 1 vs. Type 2 - Diff Location & 0.57 & 20.55 \\
    & & Type 1 vs. Type 3 - Diff Location & 0.73 & 26.41 \\
    & & Type 2 vs. Type 3 - Diff Location & 0.91 & 32.75 \\
    \hline
    \multirow{6}{*}{Google} & \multirow{6}{*}{General} & Type 1 vs. Type 2 - Same Location & 0.34 & 12.41 \\
    & & Type 1 vs. Type 3 - Same Location & 0.79 & 28.49 \\
    & & Type 2 vs. Type 3 - Same Location & 0.14 & 5.06 \\
    & & Type 1 vs. Type 2 - Diff Location & \textbf{0.02} & 0.76 \\
    & & Type 1 vs. Type 3 - Diff Location & \textbf{0.0036} & 0.13 \\
    & & Type 2 vs. Type 3 - Diff Location & 0.85 & 30.60 \\
    \hline
    \multirow{6}{*}{Yahoo} & \multirow{6}{*}{General} & Type 1 vs. Type 2 - Same Location & 0.52 & 18.74 \\
    & & Type 1 vs. Type 3 - Same Location & \textbf{0.0013} & \textbf{0.05} \\
    & & Type 2 vs. Type 3 - Same Location & \textbf{0.0036} & 0.13  \\
    & & Type 1 vs. Type 2 - Diff Location & 0.68 & 24.39 \\
    & & Type 1 vs. Type 3 - Diff Location & \textbf{0.02} & 0.62 \\
    & & Type 2 vs. Type 3 - Diff Location & \textbf{0.03} & 1.12 \\
    \hline \hline
    \multirow{6}{*}{DuckDuckGo} & \multirow{6}{*}{Specific} & Type 1 vs. Type 2 - Same Location & 0.34 & 12.41 \\
    & & Type 1 vs. Type 3 - Same Location & 0.43 & 15.38 \\
    & & Type 2 vs. Type 3 - Same Location & 0.85 & 30.60 \\
    & & Type 1 vs. Type 2 - Diff Location & \textbf{0.0073} & 0.26 \\
    & & Type 1 vs. Type 3 - Diff Location & \textbf{0.0091} & 0.33 \\
    & & Type 2 vs. Type 3 - Diff Location & 0.21 & 7.64 \\
    \hline
    \multirow{6}{*}{Google} & \multirow{6}{*}{Specific} & Type 1 vs. Type 2 - Same Location & 0.38 & 13.85 \\
    & & Type 1 vs. Type 3 - Same Location & \textbf{0.01} & 0.41 \\
    & & Type 2 vs. Type 3 - Same Location & 0.14 & 5.06 \\
    & & Type 1 vs. Type 2 - Diff Location & \textbf{0.0002} & \textbf{0.01} \\
    & & Type 1 vs. Type 3 - Diff Location & \textbf{0.0002} & \textbf{0.01} \\
    & & Type 2 vs. Type 3 - Diff Location & 0.91 & 32.75 \\
    \hline
    \multirow{6}{*}{Yahoo} & \multirow{6}{*}{Specific} & Type 1 vs. Type 2 - Same Location & 0.82 & 29.54 \\
    & & Type 1 vs. Type 3 - Same Location & \textbf{0.0036} & 0.13 \\
    & & Type 2 vs. Type 3 - Same Location & \textbf{0.0028} & 0.10 \\
    & & Type 1 vs. Type 2 - Diff Location & 0.79 & 28.49 \\
    & & Type 1 vs. Type 3 - Diff Location & \textbf{0.0091} & 0.33 \\
    & & Type 2 vs. Type 3 - Diff Location & \textbf{0.0017} & 0.06 \\
    \hline
    \end{tabular}
\end{table*}

\clearpage
\begin{table}[hb]
\caption{ANOVA statistical results of RBO values for different browsing histories comparisons with the stateless profiles, across search engines and query groups.}
\centering
\begin{tabular}{p{1cm} p{1cm} p{2cm} p{1.3cm} p{1.2cm}}
\hline
Search\newline Engine &  Query\newline Group & Brow.profiles\newline comparisons & Mean \newline D=1-RBO & ANOVA \newline Result \\
\hline
\multirow{6}{*}{Duckgo} 
& \multirow{3}{*}{\small{general}} & \small{conflict v. stat.} & 0.22 & F=1.792\\
& & general v. stat. & 0.288047 & p=0.170 \\
& & stateless v. stat. & 0.221987 &\\
\cline{2-5} 
& \multirow{3}{*}{\small{specific}} & \small{conflict v. stat.} & 0.30 & F=2.33 \\
& & \small{general v. stat.} & 0.37 & p=0.10\\
& & \small{stateless v. stat.} & 0.27&\\
\hline
\multirow{6}{*}{Google} 
& \multirow{3}{*}{\small{general}} & \small{conflict v. stat.} & 0.15  & F=3.25 \\
& & \small{general v. stat.} & 0.13& p=0.04\\
& & \small{stateless v. stat.} & 0.09 &\\
\cline{2-5}
& \multirow{3}{*}{\small{specific}} & conflict v. stat. & 0.244388 & F=8.078 \\
& & \small{general v. stat.} & 0.22 & p =0.0\\
& & \small{stateless v. stat.} & 0.12 &\\
\hline
\multirow{6}{*}{Yahoo} 
& \multirow{3}{*}{\small{general}} & \small{conflict v. stat.} & 0.29& F =0.04 \\
& & \small{general v. stat.} & 0.29 & p =0.96\\
& & \small{stateless v. stat.} & 0.28 &\\
\cline{2-5}
& \multirow{3}{*}{\small{specific}} & \small{conflict v. stat.} & 0.32 & F =0.09 \\
& & \small{general v. stat.} & 0.31 & p =0.91\\
& & \small{stateless v. stat.} & 0.33 &\\
\hline
\end{tabular}

\label{table:stat_results}
\end{table}
\clearpage

\subsection{Additional metrics of results differences}

\begin{figure}[ht]
    \centering
    \includegraphics[width=1\columnwidth]{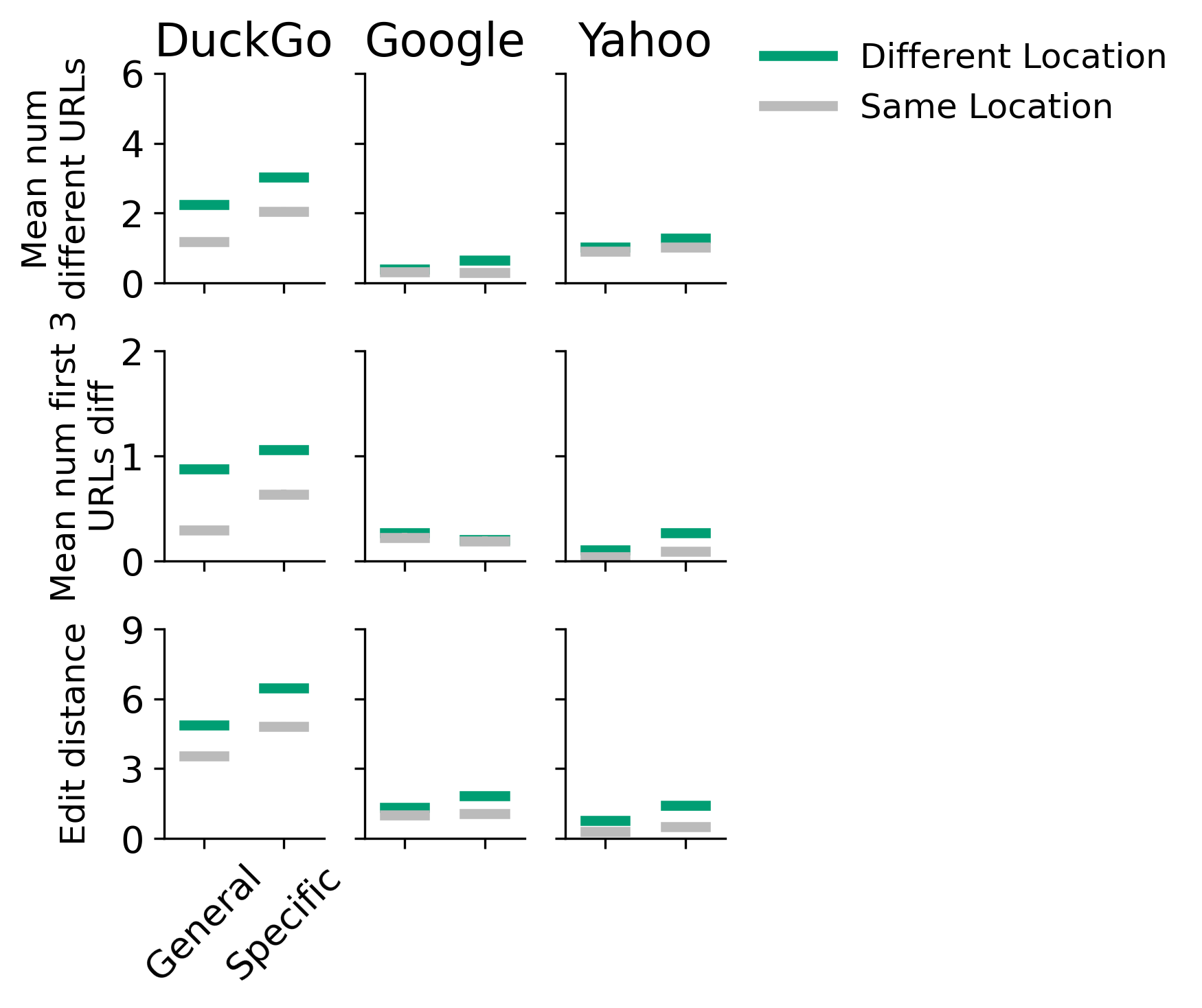}
    \caption{Results for Type 1 bots. (Top Row) Average number of URLs that were not present in both lists of results in the 10 results. (Middle row) Average number of URLs that were not present in the top 3 results for both lists of results being compared. (Bottom Row) Edit distance between lists.  Error bars represent 95\% confidence intervals based on the bootstrapped distribution.}
    \Description{}
    \label{fig:other_metrics_step_1}
\end{figure}

\begin{figure}[hb]
    \centering
    \includegraphics[width=1\columnwidth]{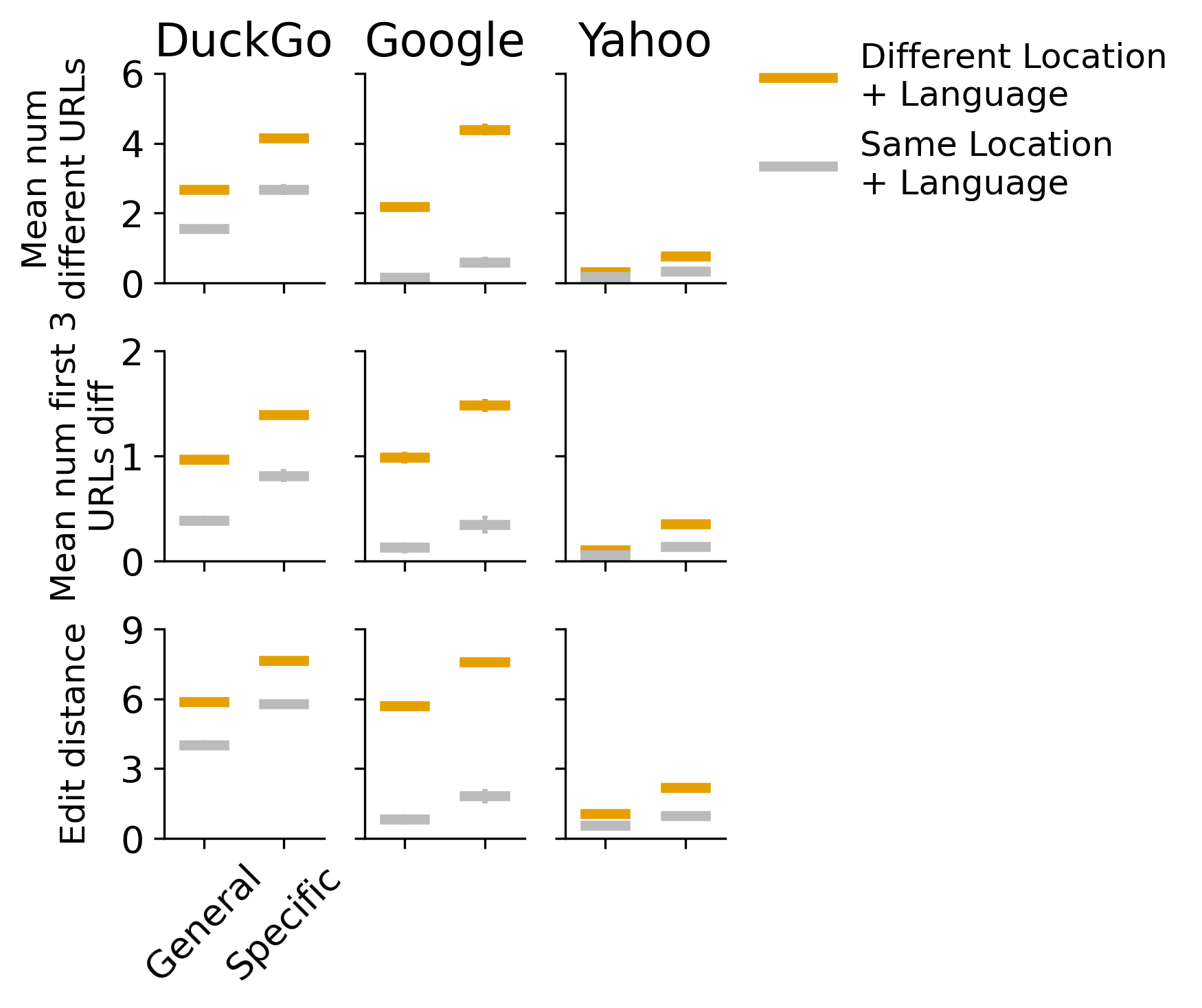}
    \caption{Results for Type 2 bots. (Top Row) Average number of URLs that were not present in both lists of results in the 10 results. (Middle row) Average number of URLs that were not present in the top 3 results for both lists of results being compared. (Bottom Row) Edit distance between lists.  Error bars represent 95\% confidence intervals based on the bootstrapped distribution.}
    \Description{}
    \label{fig:other_metrics_step_2}
\end{figure}

\begin{figure}[h]
    \centering
    \includegraphics[width=1\columnwidth]{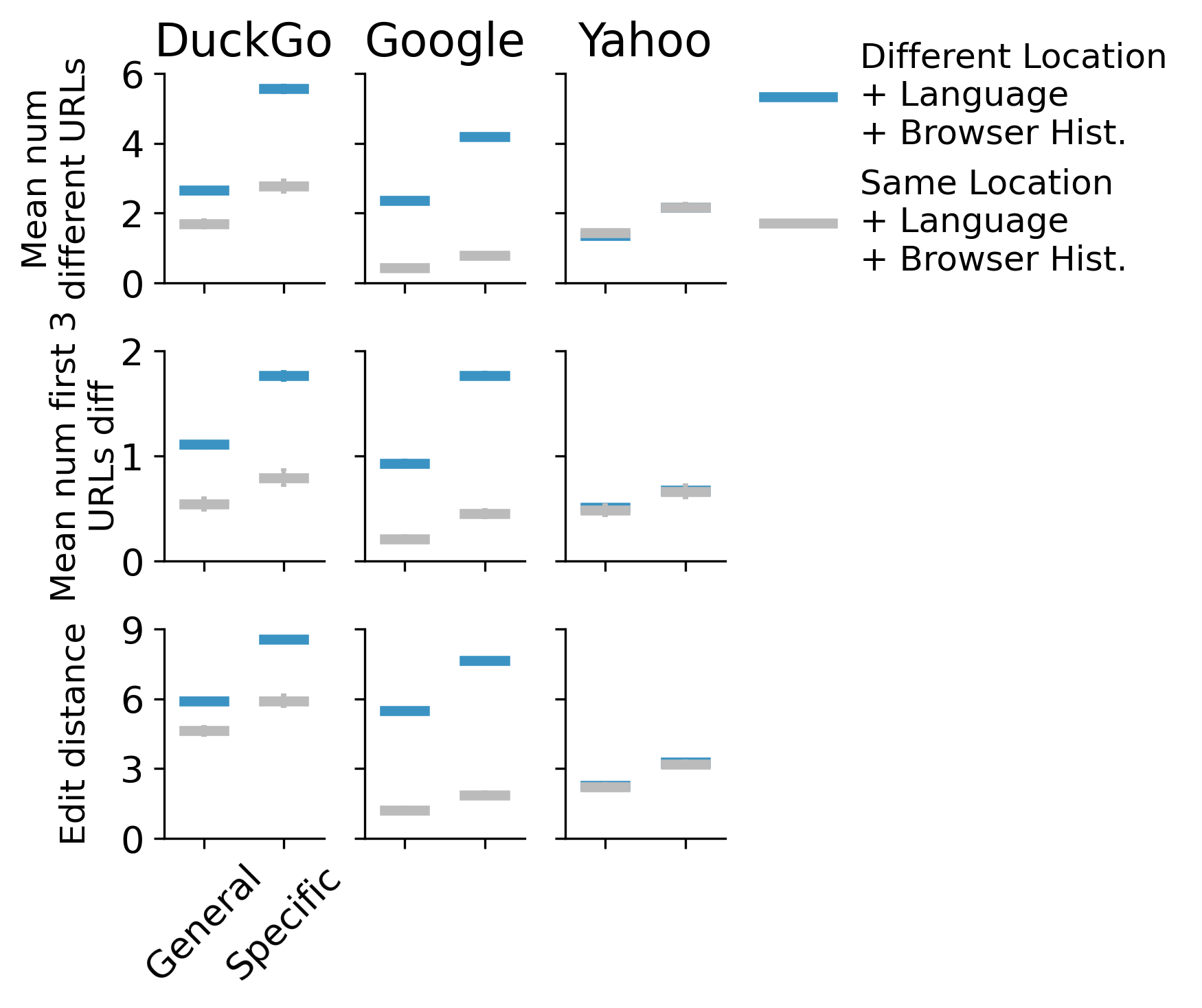}
    \caption{Results for Type 3 bots. (Top Row) Average number of URLs that were not present in both lists of results in the 10 results. (Middle row) Average number of URLs that were not present in the top 3 results for both lists of results being compared. (Bottom Row) Edit distance between lists.  Error bars represent 95\% confidence intervals based on the bootstrapped distribution.}
    \Description{}
    \label{fig:other_metrics_step_3}
\end{figure}

The $D = 1 - RBO$ measurement used in the paper calculates a value between $[0, 1]$ according to the similarity of two lists and the rank at which two elements of the list match. To complement this metric we also calculated: (1) the average number of URLs that were not present in both lists of results of the two bots being compared (top row in Figures \ref{fig:other_metrics_step_1},\ref{fig:other_metrics_step_2}, \ref{fig:other_metrics_step_3}); (2) average number of URLs that were present at both top 3 results for both lists of results being compared (middle row in Figures \ref{fig:other_metrics_step_1},\ref{fig:other_metrics_step_2}, \ref{fig:other_metrics_step_3}) and (3) the average edit distance (bottom row in Figures \ref{fig:other_metrics_step_1},\ref{fig:other_metrics_step_2}, \ref{fig:other_metrics_step_3}). 

\newpage

\subsection{Controlling for query size}

\begin{figure}[h]
    \centering
    \includegraphics[width=1\columnwidth]{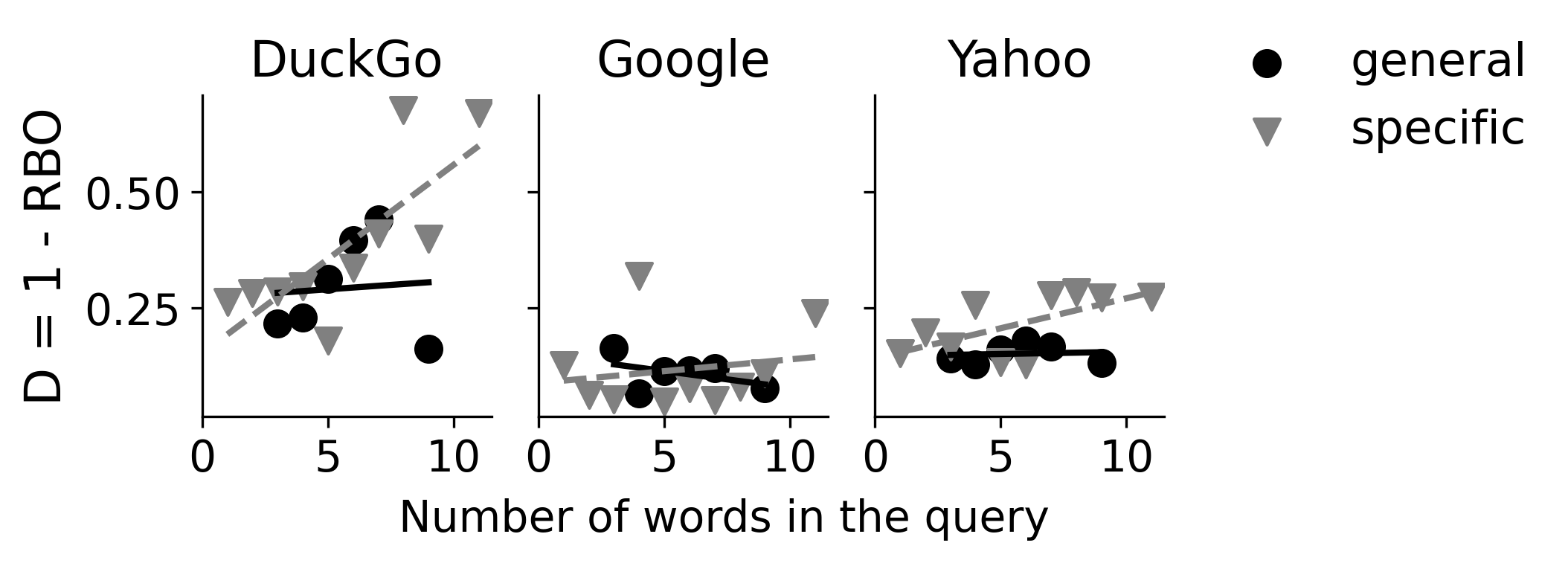}
    \caption{Average value of $D = 1 - RBO$ (Type 1 bots) by query size in terms of number of words. In black plotted with circles we have the results for queries of the general category and in grey plotted with triangles the results of queries in the specific category.}
    \Description{}
    \label{results_by_query_size}
\end{figure}

As shown in Table \ref{tab:queries_long_table} the  categories of queries have very different sizes. Therefore, we tested whether the higher values of $D = 1 - RBO$ for queries of the specific group was a consequence of a higher number of words (bigger dimension). Figure \ref{results_by_query_size} shows the value of $D = 1 - RBO$ as function of the number of words in the query for both queries of the general category (black circle) and specific category (gray triangle) across search engines. As the plot shows there is a tendency for the value of $D = 1 - RBO$ to be higher for specific queries than general ones even when comparing queries of the exact same size from both categories. 

\begin{figure}[h]
    \centering
    \includegraphics[width=1\columnwidth]{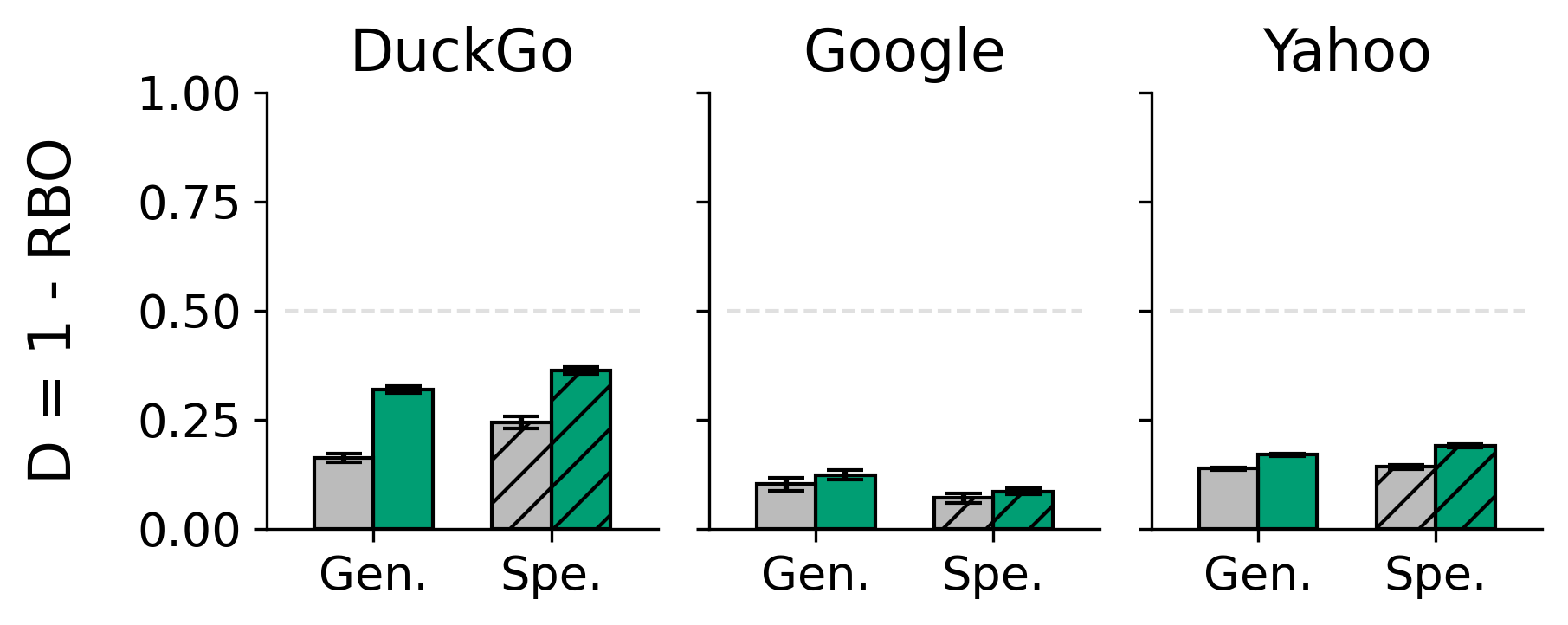}
    \caption{Average value of $D = 1 - RBO$ for bots in the same location (grey) and in different locations (green) considering exclusively queries of the same length (between 3 and 8 words).}
    \Description{}
    \label{rbo_queries_same_size}
\end{figure}

Additionally, when we compare the value of $D = 1- RBO$ per location (different and same location) for an equal number of queries of both categories with comparable sizes (within 3 and 9 words) we observe (Figure \ref{rbo_queries_same_size} that the results continue to present the same pattern as in Figure \ref{fig:main_plot}. That is, higher values of $D = 1- RBO$ for specific queries than general queries and also higher values when comparing results of bots from different locations than the same location. 

\subsection{D = 1 - RBO values for the different experimental steps (for the 10 common queries across bot types)}

\begin{figure}[h]
    \centering
    \includegraphics[width=1\columnwidth]{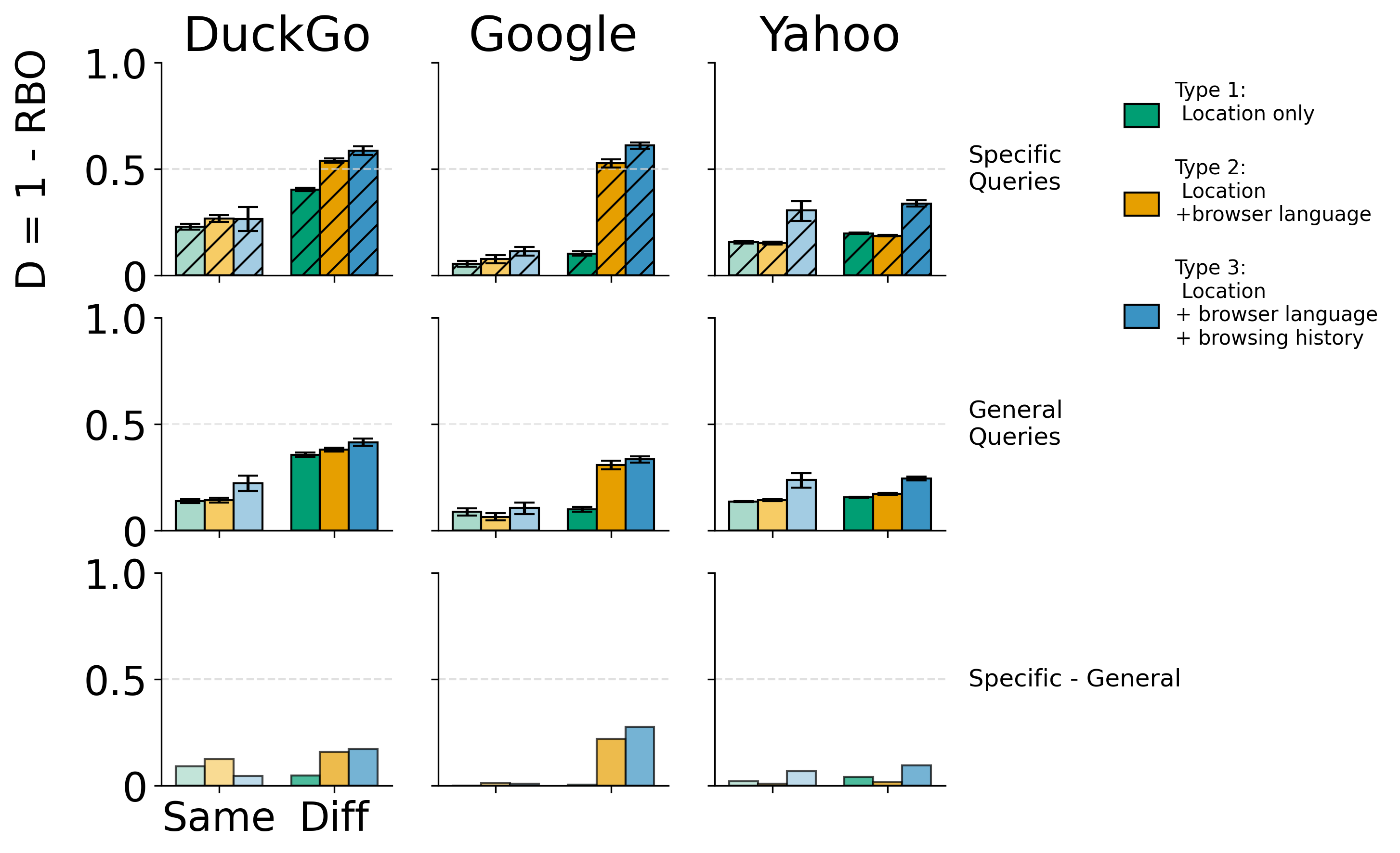}
    \caption{$D = 1 - RBO$ Results: Comparison of the experimental outcomes across search engines for the same location (left) and different locations (right). The top row illustrates the results for specific queries, the middle row shows results for general queries, and the bottom row highlights the difference between specific and general queries.}
    \label{fig:comparison_steps}
    \Description{}
\end{figure}

\subsection{Differences across categories of websites}

\begin{figure}[hb]
    \includegraphics[width=1\columnwidth]{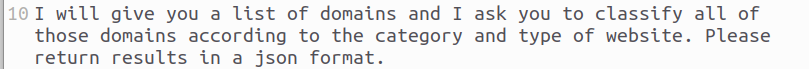}
    \caption{Prompt given to ChatGPT for the domains classification.}
    \Description{}
    \label{prompt_domain_cat}
\end{figure}

To control for the possibility that bots were retrieving the same content but in localized versions, leading to differences in URLs, we used ChatGPT-4o to classify the websites by category and type. The prompt was the one depicted in Figure \ref{prompt_domain_cat}.

\begin{figure}[ht]
    \centering
    \includegraphics[width=1\columnwidth]{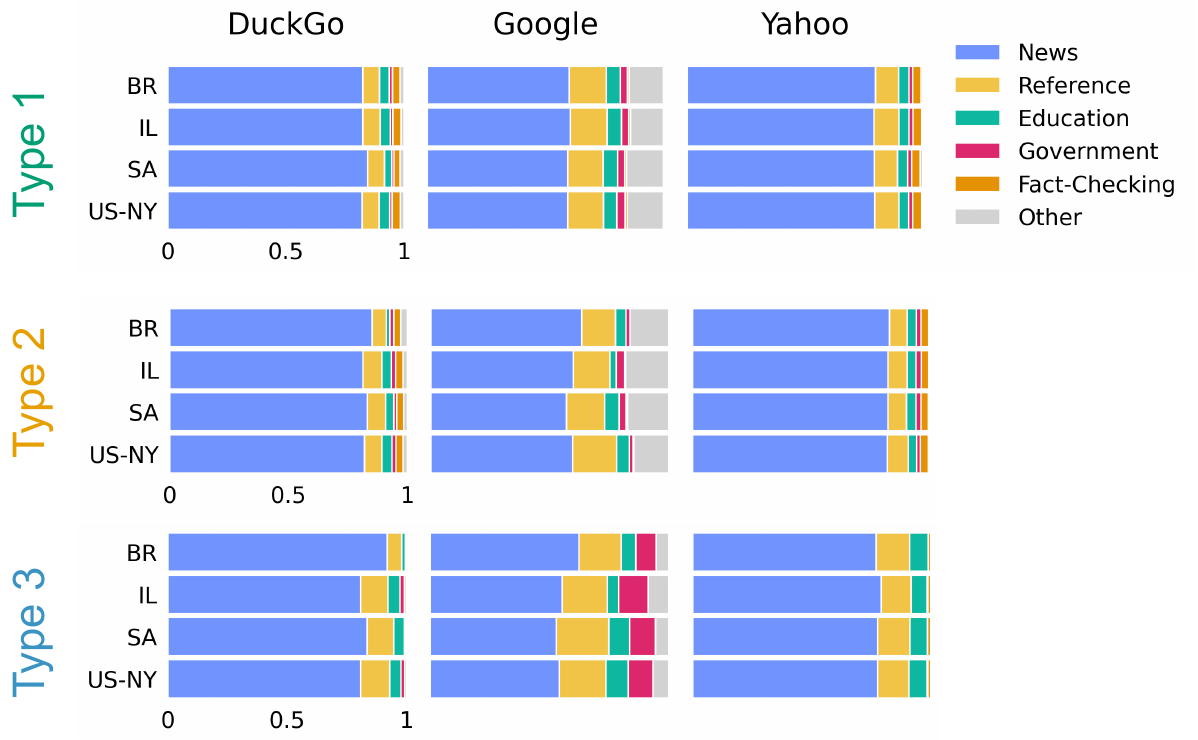}
    \caption{Proportion of results per top 5 category of website for the specific group of queries, per search engine, country and bot type.}
    \label{categories_websites_specific}
    \Description{}
\end{figure}

\begin{figure}[hb]
    \centering
    \includegraphics[width=1\columnwidth]{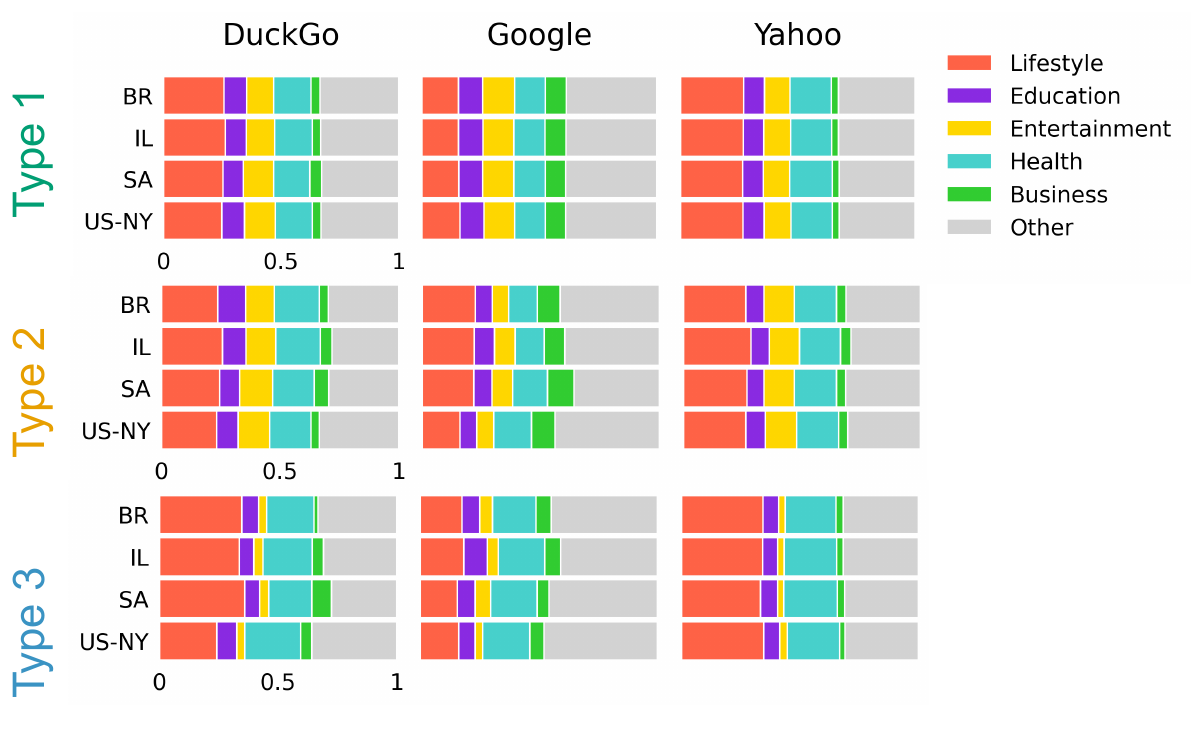}
    \caption{Proportion of results per top 5 category of website for the general group of queries, per search engine, country and bot type.}
    \label{categories_websites_general}
    \Description{}
\end{figure}

All categories per query type are in Table \ref{tab:categories_category}. Figure \ref{categories_websites_general}, shows the most common categories across search engines and location, for different Type of bots for general queries. Figure \ref{categories_websites_specific} refers to specific queries. 

\begin{table}[h]
    \centering
    \caption{Classifications of ChatGPT for category of results websites associated with General and Specific queries.}
    \label{tab:categories_category}
    \begin{tabular}{|p{3cm}|p{4cm}|}
    \hline
    \textbf{Query category} & \textbf{Websites classifications}\\
    \hline
    General & Reference, Entertainment, Education, Technology, News, Lifestyle, Business, Finance, Health, Government, Non-Profit, Social Media, Travel, E-Commerce, Art, Science, Fashion, Legal, Career, Retail, Automotive, Food \\
    \hline
    Specific & Reference, Education, Government, News, Fact-Checking, Social Media, Non-Profit, Entertainment, Finance, Religion, E-Commerce, Technology, Sports, Travel, Science \\
    \hline
    \end{tabular}
\end{table}

Figure \ref{categories_websites_specific} and \ref{categories_websites_general}reveal noticeable differences in the distribution of categories between the different query groups. 

% \begin{table}[h!]
%     \centering
%     \caption{Average number of websites categories per experimental step and search engine for both categories of queries (General and Specific).}
%     \label{tab:number_categories}
%     \begin{tabular}{|p{0.8cm}|p{2cm}|p{1.5cm}|p{1.5cm}|}
%     \hline
%     \textbf{} & \textbf{Search Engine} & \textbf{General Queries} & \textbf{Specific Queries}\\
%     \hline
%     \multirow{4}{*}{Step 1} & Bing & 2.90 & 2.19 \\
%     \cline{2-4}
%     {} & DuckDuckGo & 3.09 & 2.19 \\
%     \cline{2-4}
%     {} & Google & 4.01 & 3.39 \\
%     \cline{2-4}
%     {} & Yahoo & 2.54 & 1.96  \\
%     \hline
%     \multirow{4}{*}{Step 2} & Bing & 2.90 & 2.49 \\
%     \cline{2-4}
%     {} & DuckDuckGo & 2.92 & 2.15 \\
%     \cline{2-4}
%     {} & Google & 3.97 & 3.42\\
%     \cline{2-4}
%     {} & Yahoo & 2.59 & 1.86 \\
%     \hline
%     \multirow{4}{*}{Step 3} & Bing & 2.67 &  2.29 \\
%     \cline{2-4}
%     {} & DuckDuckGo & 3.15 & 1.77 \\
%     \cline{2-4}
%     {} & Google & 4.16 & 3.20\\
%     \cline{2-4}
%     {} & Yahoo & 2.75 & 1.68\\
%     \hline
%     \end{tabular}
% \end{table}

\begin{figure}[ht]
    \centering
    \includegraphics[width=1\columnwidth]{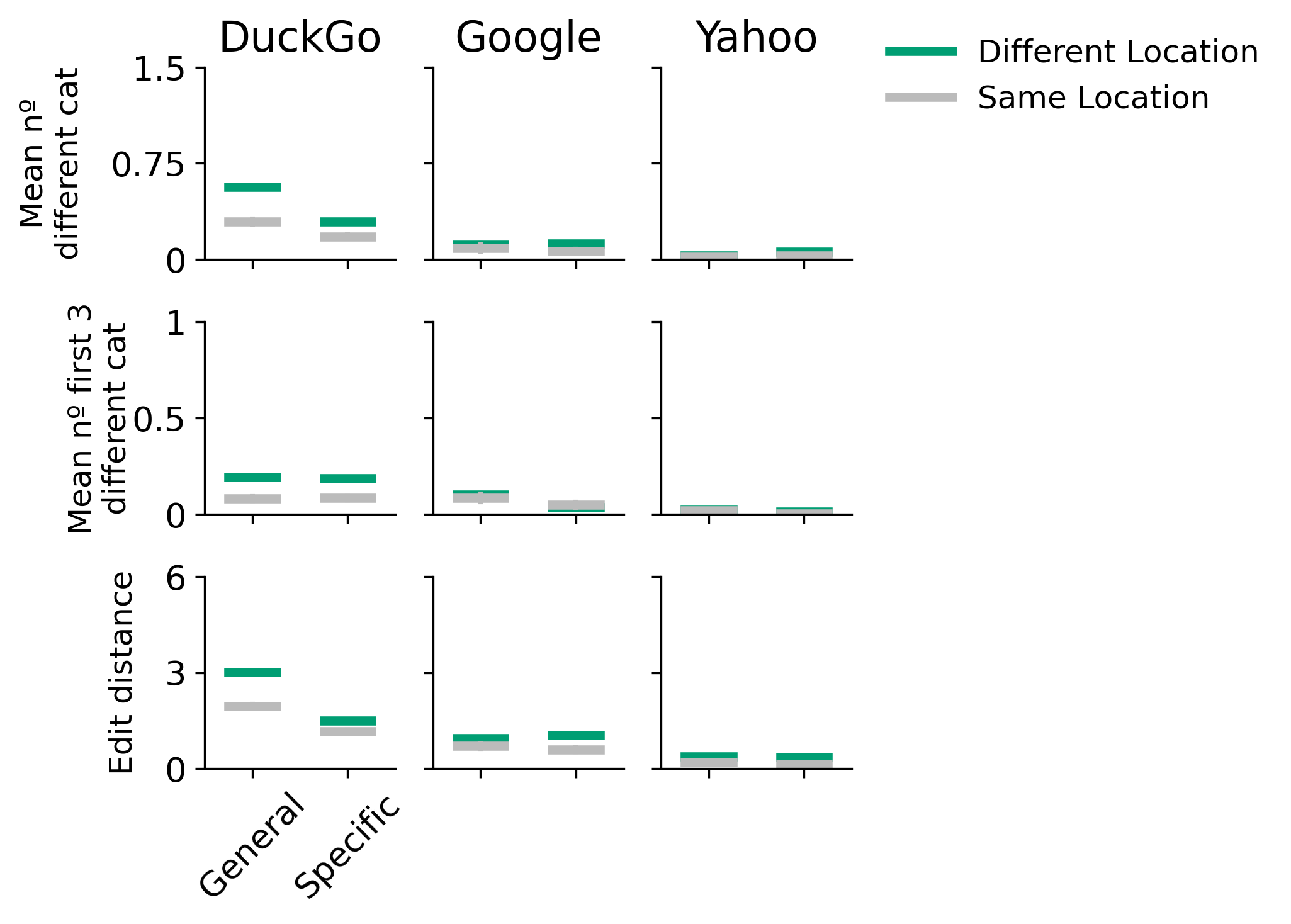}
    \caption{Results for Type 1 bots. (Top Row) Average number of categories that were not present in both lists of results in the 10 results. (Middle row) Average number of categories that were not present in the top 3 results for both lists of results being compared. (Bottom Row) Edit distance between lists of categories.  Error bars represent 95\% confidence intervals based on the bootstrapped distribution.}
    \Description{}
    \label{fig:other_categories_step_1}
\end{figure}

\begin{figure}[hb]
    \centering
    \includegraphics[width=1\columnwidth]{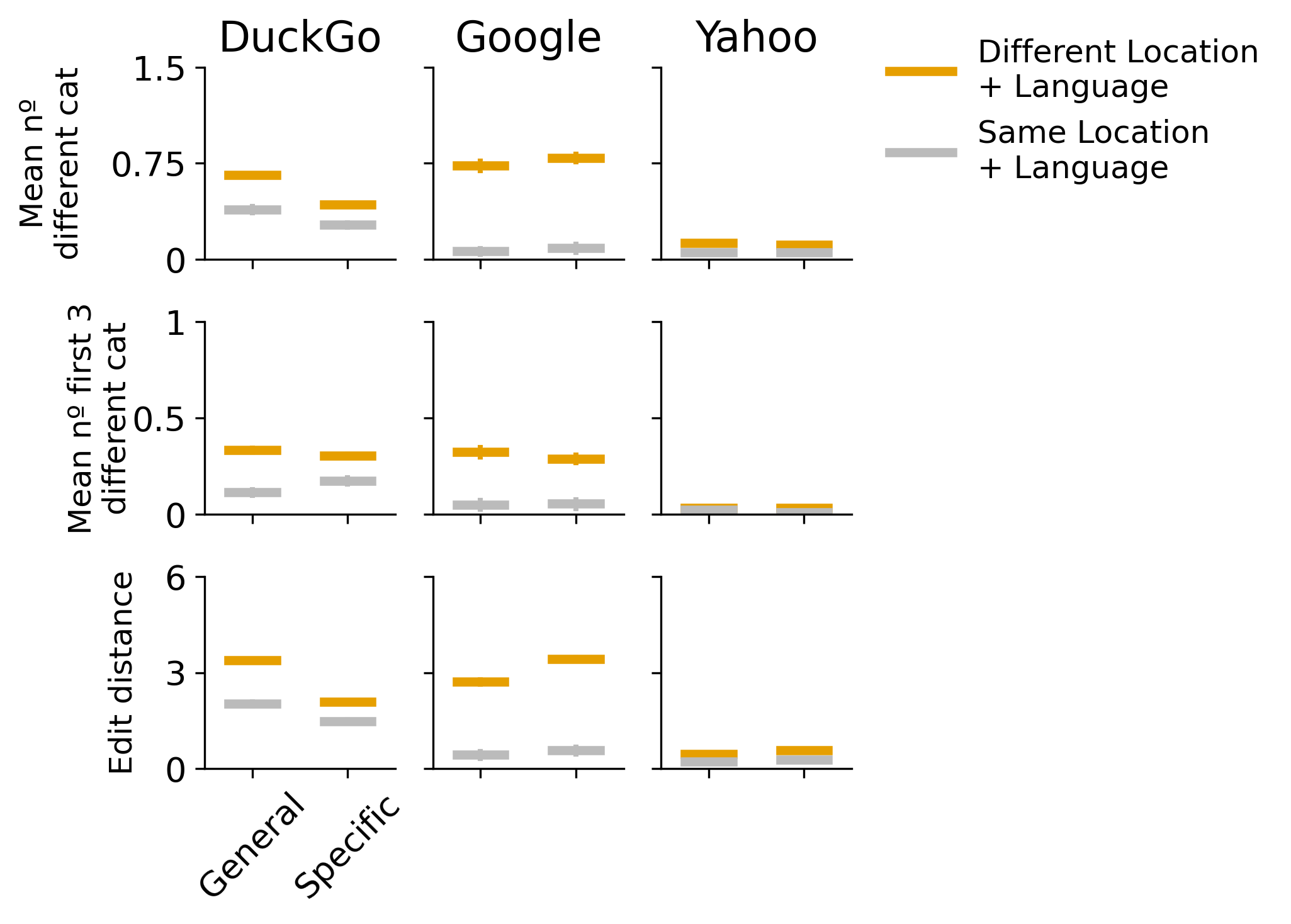}
    \caption{Results for Type 2 bots. (Top Row) Average number of categories that were not present in both lists of results in the 10 results. (Middle row) Average number of categories that were not present in the top 3 results for both lists of results being compared. (Bottom Row) Edit distance between lists of categories.  Error bars represent 95\% confidence intervals based on the bootstrapped distribution.}
    \Description{}
    \label{fig:other_categories_step_2}
\end{figure}

\begin{figure}[hb]
    \centering
    \includegraphics[width=1\columnwidth]{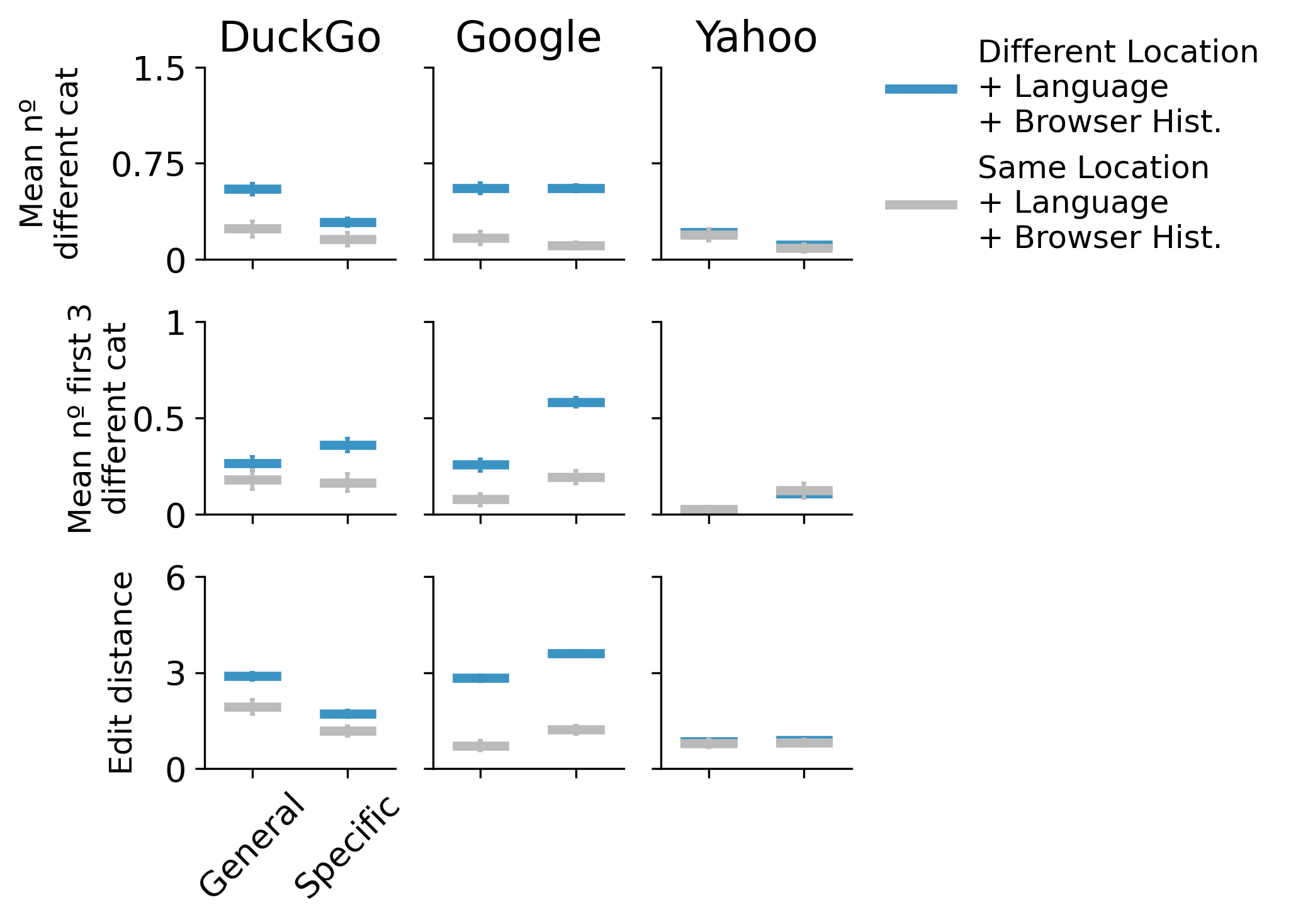}
    \caption{Results for Type 3 bots. (Top Row) Average number of categories that were not present in both lists of results in the 10 results. (Middle row) Average number of categories that were not present in the top 3 results for both lists of results being compared. (Bottom Row) Edit distance between lists of categories.  Error bars represent 95\% confidence intervals based on the bootstrapped distribution.}
    \Description{}
    \label{fig:other_categories_step_3}
\end{figure}

After classifying all domains present in all experimental steps (Table \ref{tab:categories_category}) we studied how the results varied as function of the location and query type for all bots types. Additionally to the metric $D = 1 - RBO$ we also calculated (1) the average number of categories that were not present in both lists of results of the two bots being compared (top row in Figures \ref{fig:other_categories_step_1},\ref{fig:other_categories_step_2}, \ref{fig:other_categories_step_3}); (2) average number of URLs that were present at both top 3 results for both lists of results being compared (middle row in Figures \ref{fig:other_categories_step_1},\ref{fig:other_categories_step_2}, \ref{fig:other_categories_step_3})) and (3) the average edit distance (bottom row in Figures \ref{fig:other_categories_step_1},\ref{fig:other_categories_step_2}, \ref{fig:other_categories_step_3})). 

Additionally, we show in \ref{tab:p_values_adjusted_categories_step_1}, \ref{tab:p_values_adjusted_categories_step_2} and \ref{tab:p_values_adjusted_categories_step_3} more information about the statistical values plotted in \ref{fig:rbo_per_category_website}. 
\newpage

\begin{table}[hb]
    \centering
    \caption{Pairwise comparisons of RBO values for categories of websites for Type 1 bots and each search engine, showing p-values and Bonferroni-adjusted p-values. The Mann-Whitney U test was used for the comparisons, and significant adjusted p-values ($<0.05$) are highlighted in bold. The adjustment accounts for 12 comparisons across the three search engines.}
    \Description{}
    \label{tab:p_values_adjusted_categories_step_1}
    \begin{tabular}{|p{1cm}|p{3cm}|p{1cm}|p{1.5cm}|}
    \hline
    \textbf{Search\newline Engine} & \textbf{Groups \newline compared} & \textbf{p-value} & \textbf{Adjusted \newline p-value} \\
    \hline\hline
    \multirow{4}{*}{DuckGo} & Same Location vs. Diff Location - General Queries & \textbf{0.0014} & \textbf{0.0014} \\
    \cline{2-4}
    & Same Location vs. Diff Location - Specific Queries & 0.0659 & 0.7905 \\
    \cline{2-4}
    & General vs. Specific - Same Location & 0.1959 & 2.3512 \\
    \cline{2-4}
    & General vs. Specific - Diff Location & 0.2866 & 3.4392 \\
    \hline \hline
    \multirow{4}{*}{Google} & Same Location vs. Diff Location - General Queries & 0.5375 & 6.4501 \\
    \cline{2-4}
    & Same Location vs. Diff Location - Specific Queries & 0.3560 & 4.2714 \\
    \cline{2-4}
    & General vs. Specific - Same Location & 0.6808 & 8.1692 \\
    \cline{2-4}
    & General vs. Specific - Diff Location & 0.6895 & 8.2741 \\
    \hline \hline
    \multirow{4}{*}{Yahoo} & Same Location vs. Diff Location - General Queries & 0.8202 & 9.8419 \\
    \cline{2-4}
    & Same Location vs. Diff Location - Specific Queries & 0.2639 & 3.1672 \\
    \cline{2-4}
    & General vs. Specific - Same Location & 0.7807 & 9.3680 \\
    \cline{2-4}
    & General vs. Specific - Diff Location & 0.5738 & 6.8860 \\
    \hline
    \end{tabular}
\end{table}
\clearpage
\begin{table}[hb]
    \centering
    \caption{Pairwise comparisons of RBO values for categories of websites for Type 2 bots and each search engine, showing p-values and Bonferroni-adjusted p-values. The Mann-Whitney U test was used for the comparisons, and significant adjusted p-values ($<0.05$) are highlighted in bold. The adjustment accounts for 12 comparisons across the three search engines.}
    \Description{}
    \label{tab:p_values_adjusted_categories_step_2}
    \begin{tabular}{|p{1cm}|p{3cm}|p{1cm}|p{1.5cm}|}
    \hline
    \textbf{Search\newline Engine} & \textbf{Groups \newline compared} & \textbf{p-value} & \textbf{Adjusted \newline p-value} \\
    \hline\hline
    \multirow{4}{*}{DuckGo} & Same Location vs. Diff Location - General Queries & \textbf{0.00} & \textbf{0.03} \\
    \cline{2-4}
    & Same Location vs. Diff Location - Specific Queries & \textbf{0.01} & \textbf{0.15} \\
    \cline{2-4}
    & General vs. Specific - Same Location & 0.81 & 9.76 \\
    \cline{2-4}
    & General vs. Specific - Diff Location & 0.73 & 8.72 \\
    \hline \hline
    \multirow{4}{*}{Google} & Same Location vs. Diff Location - General Queries & \textbf{0.00} & \textbf{0.01} \\
    \cline{2-4}
    & Same Location vs. Diff Location - Specific Queries & \textbf{0.00} & \textbf{0.00} \\
    \cline{2-4}
    & General vs. Specific - Same Location & 0.41 & 4.87 \\
    \cline{2-4}
    & General vs. Specific - Diff Location & 0.75 & 9.01 \\
    \hline \hline
    \multirow{4}{*}{Yahoo} & Same Location vs. Diff Location - General Queries & 0.66 & 7.96 \\
    \cline{2-4}
    & Same Location vs. Diff Location - Specific Queries & \textbf{0.34} & 4.09 \\
    \cline{2-4}
    & General vs. Specific - Same Location & 0.83 & 10.01 \\
    \cline{2-4}
    & General vs. Specific - Diff Location & 0.72 & 8.61 \\
    \hline
    \end{tabular}
\end{table}

\begin{table}[hb]
    \centering
    \caption{Pairwise comparisons of RBO values for categories of websites for Type 3 bots and each search engine, showing p-values and Bonferroni-adjusted p-values. The Mann-Whitney U test was used for the comparisons, and significant adjusted p-values ($<0.05$) are highlighted in bold. The adjustment accounts for 12 comparisons across the three search engines.}
    \Description{}
    \label{tab:p_values_adjusted_categories_step_3}
    \begin{tabular}{|p{1cm}|p{3cm}|p{1cm}|p{1.5cm}|}
    \hline
    \textbf{Search\newline Engine} & \textbf{Groups \newline compared} & \textbf{p-value} & \textbf{Adjusted \newline p-value} \\
    \hline\hline
    \multirow{4}{*}{DuckGo} & Same Location vs. Diff Location - General Queries & 0.5054 & 6.0643 \\
    \cline{2-4}
    & Same Location vs. Diff Location - Specific Queries & \textbf{0.0412} & 0.4941 \\
    \cline{2-4}
    & General vs. Specific - Same Location & 0.90 & 10.76 \\
    \cline{2-4}
    & General vs. Specific - Diff Location & 0.41 & 4.90 \\
    \hline \hline
    \multirow{4}{*}{Google} & Same Location vs. Diff Location - General Queries & 0.09 & 1.12 \\
    \cline{2-4}
    & Same Location vs. Diff Location - Specific Queries & \textbf{0.0004} & \textbf{0.005} \\
    \cline{2-4}
    & General vs. Specific - Same Location & \textbf{0.03} & 0.37 \\
    \cline{2-4}
    & General vs. Specific - Diff Location & \textbf{0.05} & 0.67 \\
    \hline \hline
    \multirow{4}{*}{Yahoo} & Same Location vs. Diff Location - General Queries & 0.9296 & 11.16 \\
    \cline{2-4}
    & Same Location vs. Diff Location - Specific Queries & 0.88 & 10.56 \\
    \cline{2-4}
    & General vs. Specific - Same Location & 0.87 & 10.44 \\
    \cline{2-4}
    & General vs. Specific - Diff Location & 0.71 & 8.55 \\
    \hline
    \end{tabular}
\end{table}

\clearpage
\subsection{Leaning Analysis}
\begin{figure}[h]
    \includegraphics[width=1\columnwidth]{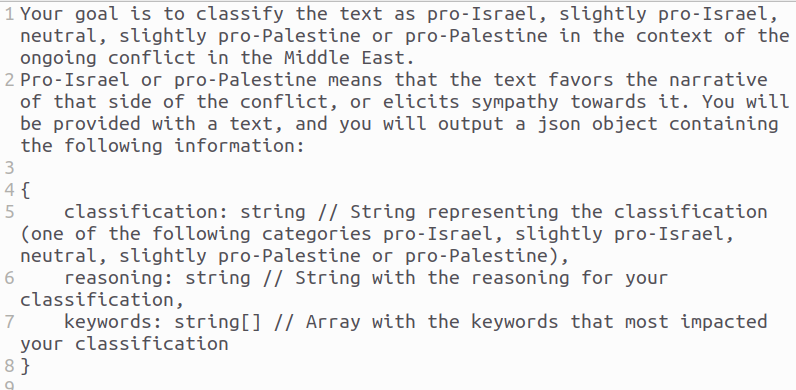}
    \caption{Prompt given to ChatGPT for classification of news leaning in the perspecitive of the Israel-Palestine conflict.}
    \Description{}
    \label{prompt_leaning}
\end{figure}

To ask ChatGPT-4o to classify the leaning of news articles, we gave ChatGPT the prompt shown in Figure \ref{prompt_leaning}.

\subsection{Mturkers Methodology}

\begin{figure}[h]
    \includegraphics[width=1\columnwidth]{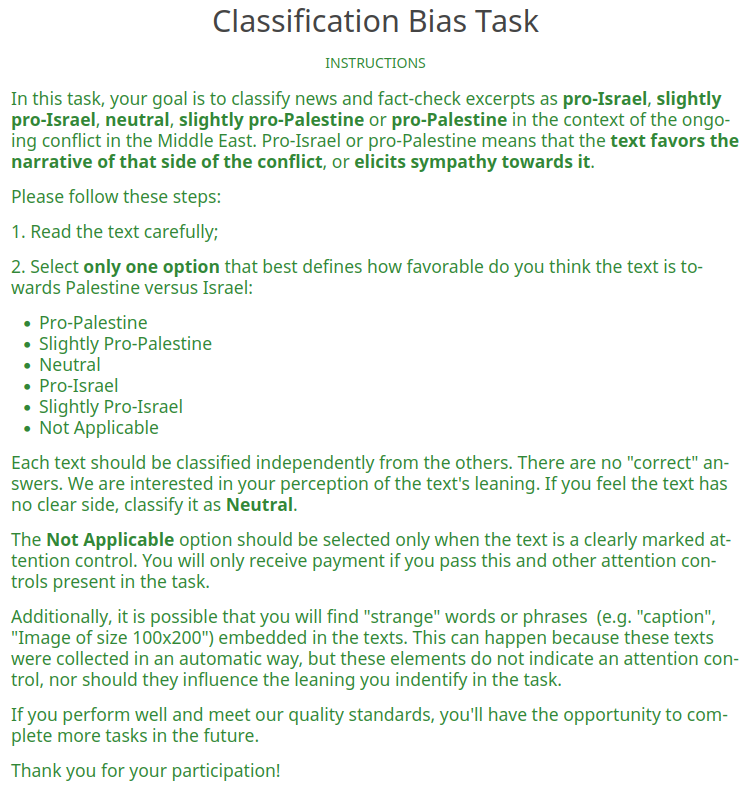}
    \caption{Instructions given to each Mturker selected to classify texts.}
    \Description{}
    \label{mturker_instructions}
\end{figure}

All Mturkers classifying text received the instructions present in Figure \ref{mturker_instructions}.

\begin{figure}[h]
    \includegraphics[width=1\columnwidth]{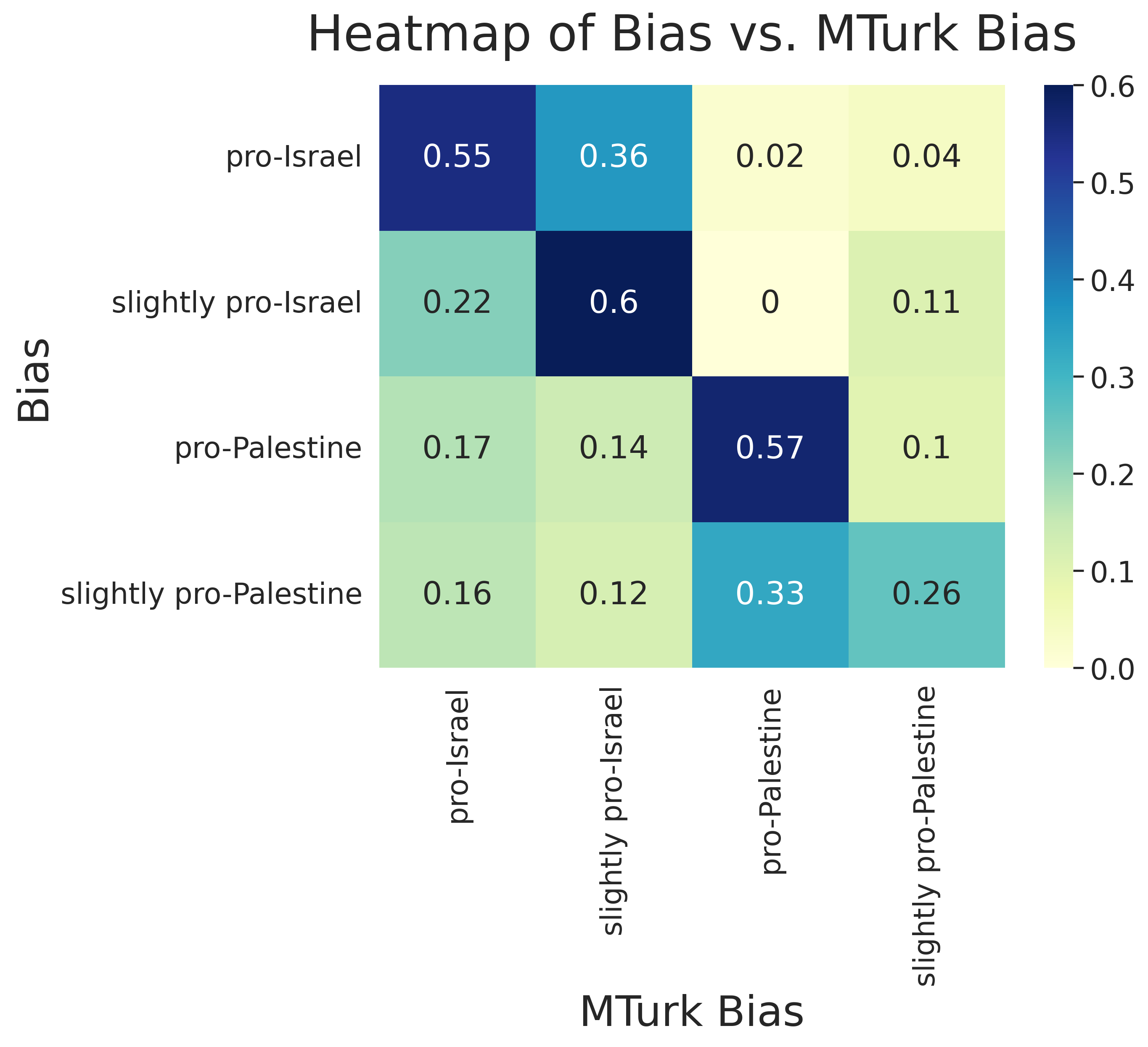}
    \caption{Heatmap displaying the proportion of bias alignments between annotators and MTurk workers, comparing different bias categories: pro-Israel, slightly pro-Israel, pro-Palestine, and slightly pro-Palestine. We excluded neutral for clarity but it is account for in the proportions. The intensity of the color indicates the proportion of agreement or alignment across the different categories.}
    \Description{}
    \label{heat_map_no_neutral}
\end{figure}

\begin{figure}[h]
    \includegraphics[width=1\columnwidth]{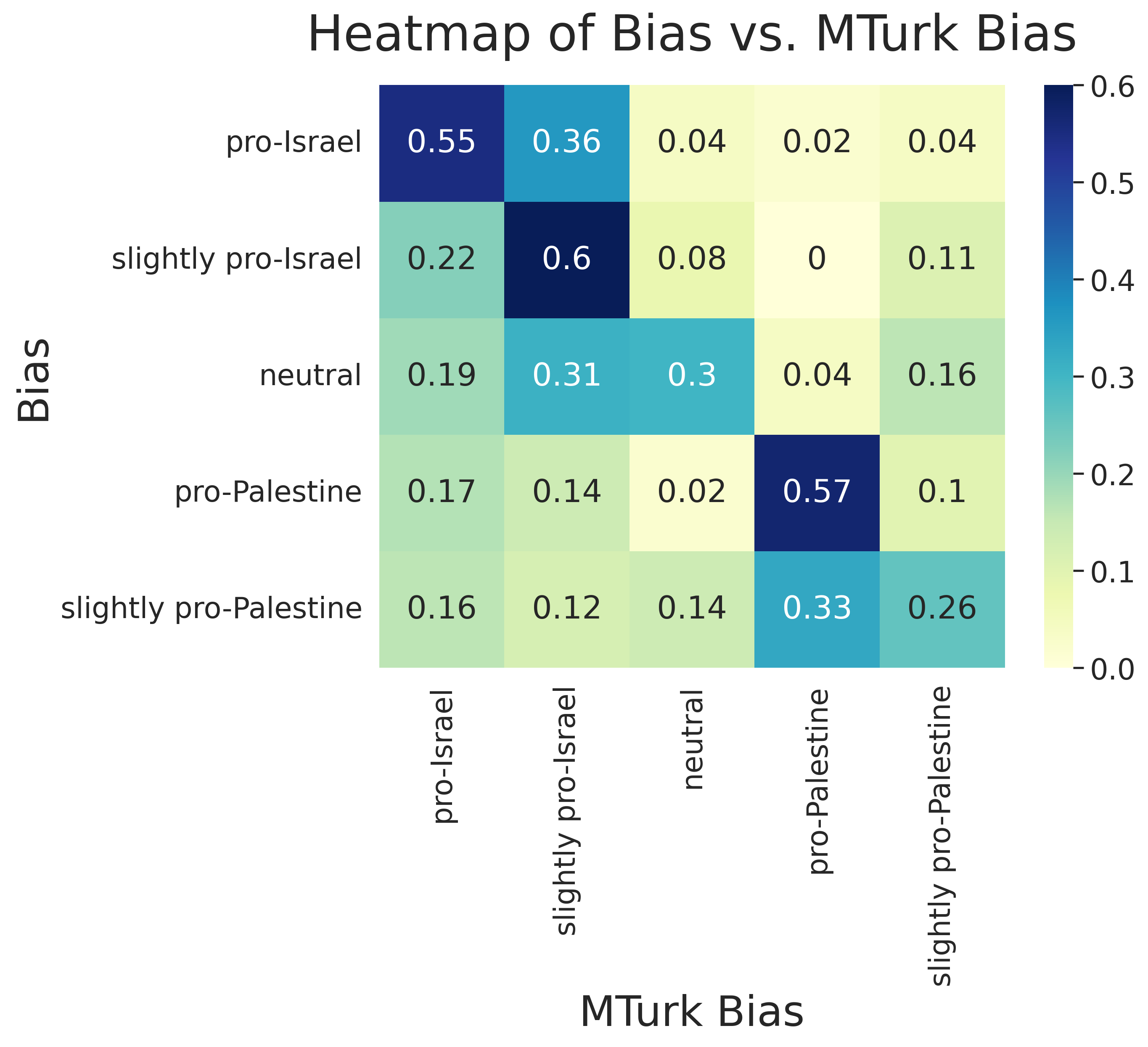}
    \caption{Heatmap displaying the proportion of bias alignments between annotators and MTurk workers, comparing different bias categories: pro-Israel, slightly pro-Israel, neutral, pro-Palestine, and slightly pro-Palestine. The intensity of the color indicates the proportion of agreement or alignment across the different categories.}
    \Description{}
    \label{heat_map}
\end{figure}

\subsection{Time control}

To measure the average variation of search engines results over time, we performed a control by deploying Type 2 webcrawlers in different moments in time (April, May and October). The goal was to understand the overtime variation of the page contents and consequent differences in results. Even if we observe some variation due to the small number of ports per location, smaller number of locations (US and SA) and smaller number of queries being compared (8 per category), we observe no specific trend, except for Yahoo where our differences were never really significant. 

\begin{figure}[h]
    \includegraphics[width=1\columnwidth]{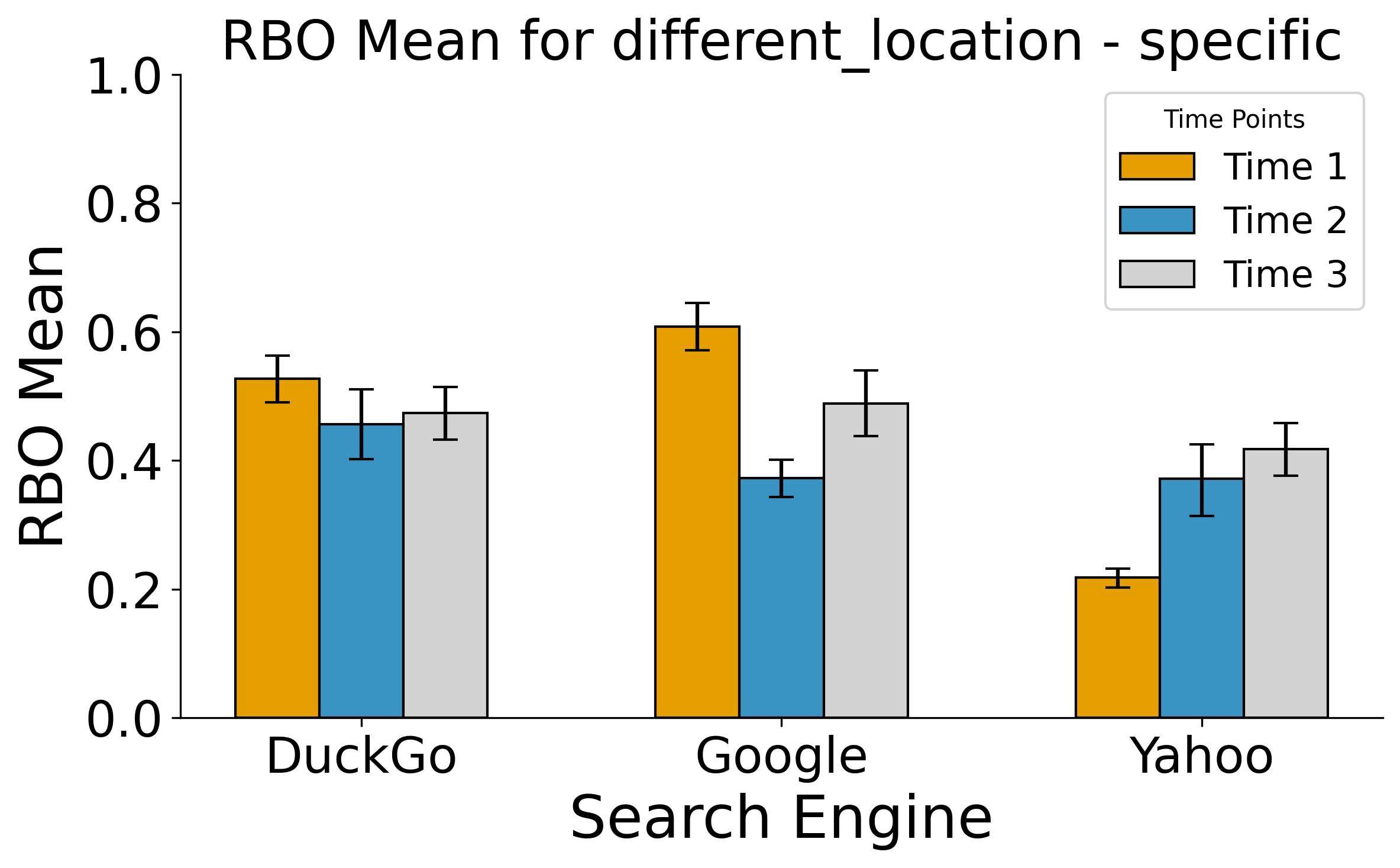}
    \caption{Average values of D = 1- RBO for the same type of bot (Type 2) with different locations (US vs. SA) across different moments in time (Time 1- April, Time 2-May and Time 3- October), for specific queries. Only 8 ports per location were considered.}
    \Description{}
    \label{time_control_1}
\end{figure}

\begin{figure}[h]
    \includegraphics[width=1\columnwidth]{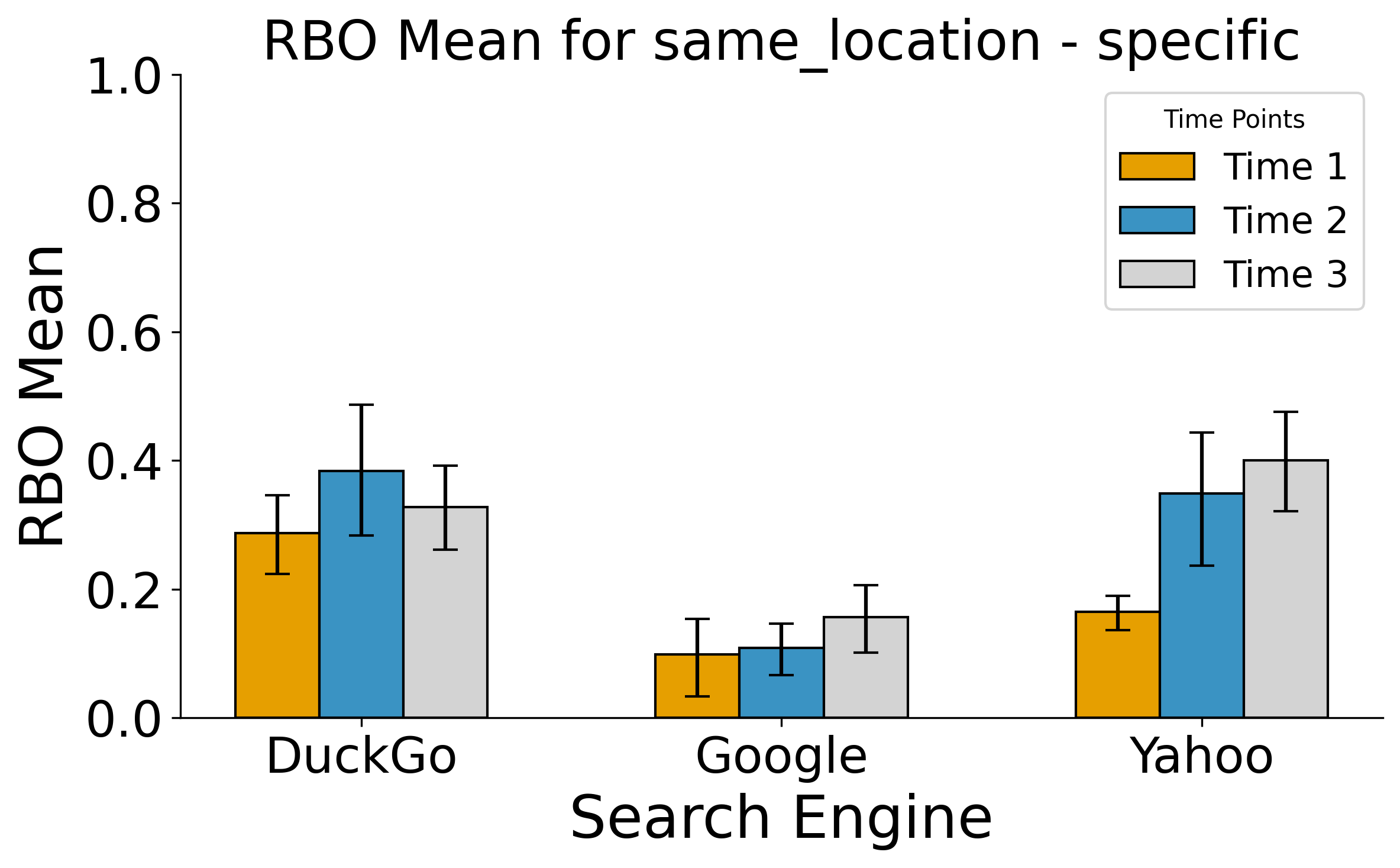}
    \caption{Average values of D = 1- RBO for the same type of bot (Type 2) with same locations (US vs. US or SA vs. SA) across different moments in time (Time 1- April, Time 2-May and Time 3- October), for specific queries. Only 8 ports per location were considered.}
    \Description{}
    \label{time_control_2}
\end{figure}

\begin{figure}[h]
    \includegraphics[width=1\columnwidth]{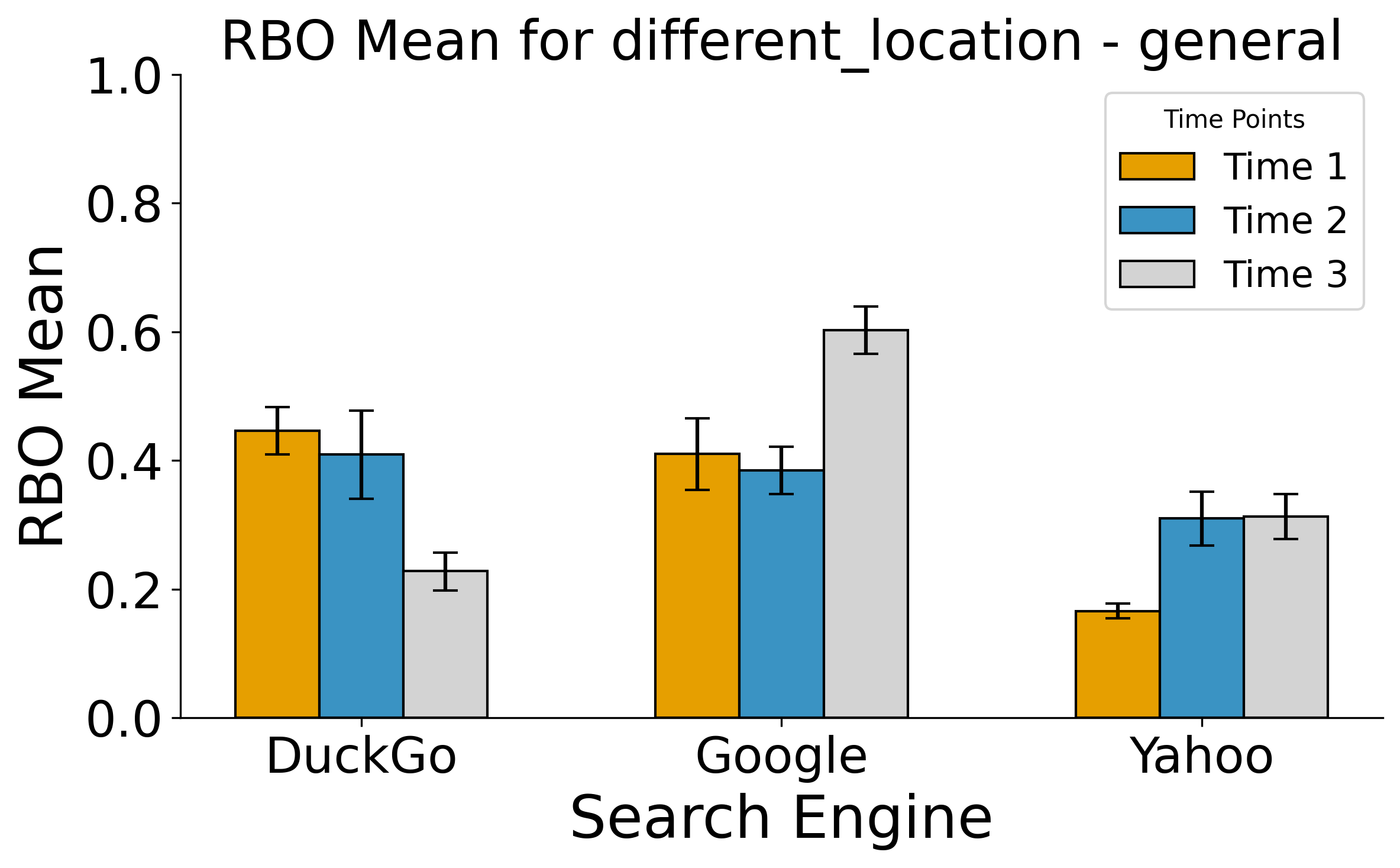}
    \caption{Average values of D = 1- RBO for the same type of bot (Type 2) with different locations (US vs. SA) across different moments in time (Time 1- April, Time 2-May and Time 3- October), for general queries. Only 8 ports per location were considered.}
    \Description{}
    \label{time_control_3}
\end{figure}

\begin{figure}[h]
    \includegraphics[width=1\columnwidth]{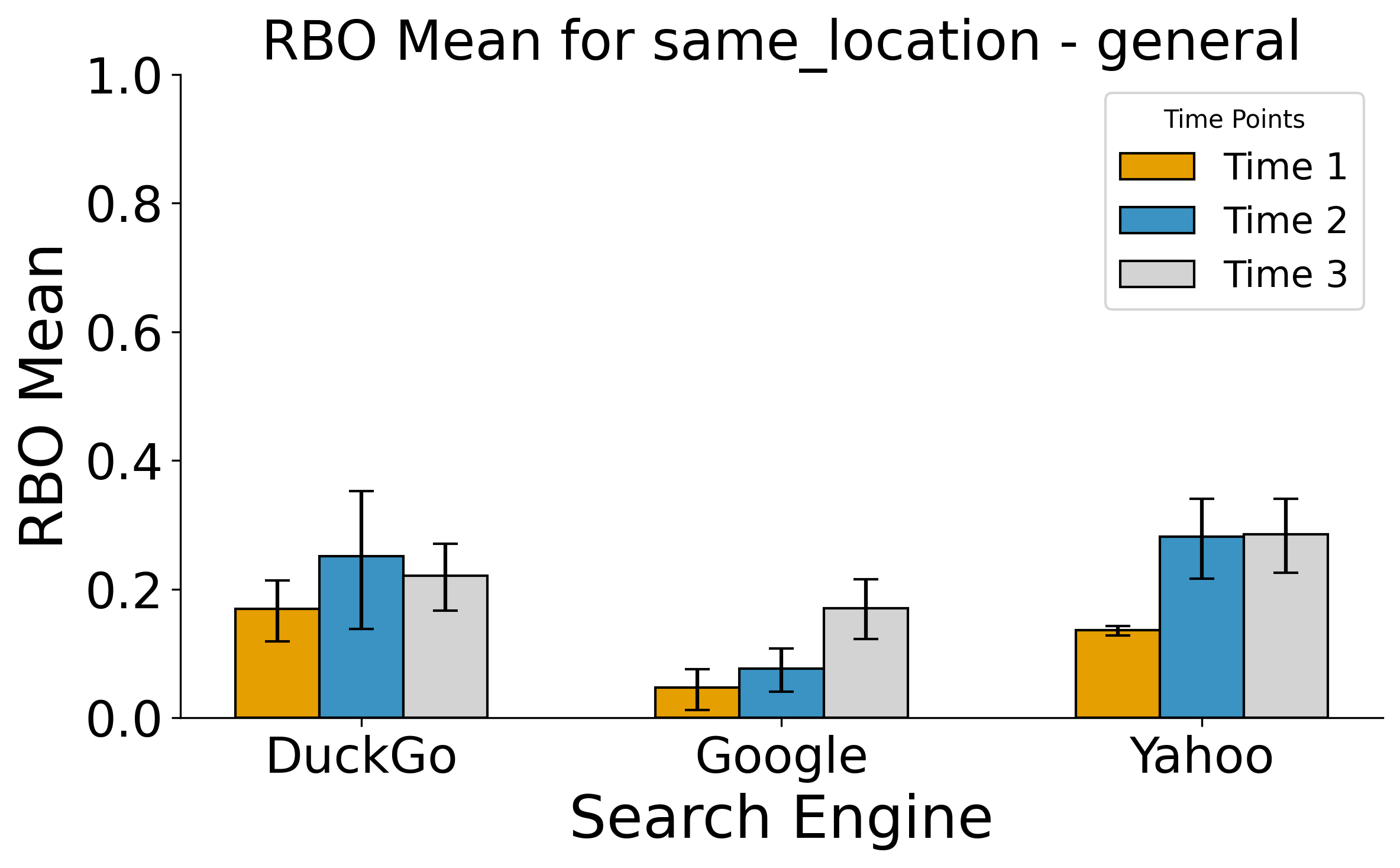}
    \caption{Average values of D = 1- RBO for the same type of bot (Type 2) with same locations (US vs. US or SA vs. SA) across different moments in time (Time 1- April, Time 2-May and Time 3- October), for general queries. Only 8 ports per location were considered.}
    \Description{}
    \label{time_control_4}
\end{figure}

% \begin{figure}[h!]
%     \centering
%     \includegraphics[width=0.8\columnwidth]{Figures/venn_unique_news_controled_step_3.png}
%     \caption{Venn diagram of unique domains present in Step 1 - Location only (green), Step 2 - Location + browser languages (yellow) and Step 3 - Location + browser languages + browsing history (blue) and its intersections.}
%     \Description{}
%     \label{venn_unique_news}
% \end{figure}

% \begin{table}[h!]
%     \centering
%     \caption{Number of unique news domains per search engine.}
%     \label{tab:unique_news}
%     \begin{tabular}{|p{0.8cm}|p{2cm}|p{2cm}|}
%     \hline
%     {} & \textbf{Search Engine} & \textbf{Num Unique News Websites} \\
%     \hline
%     \multirow{4}{*}{Step 1} & Bing &  44 \\
%     \cline{2-3}
%     {} & Duckduckgo & 39 \\
%     \cline{2-3}
%     {} & Google &  26 \\
%     \cline{2-3}
%     {} & Yahoo & 20 \\
%     \hline
%     \multirow{4}{*}{Step 2} & Bing & 49 \\
%     \cline{2-3}
%     {} & Duckduckgo & 47 \\
%     \cline{2-3}
%     {} & Google & 34 \\
%     \cline{2-3}
%     {} & Yahoo &  21 \\
%     \hline
%     \multirow{4}{*}{Step 3} &  Bing & 55 \\
%     \cline{2-3}
%     {} & Duckduckgo & 51 \\
%     \cline{2-3}
%     {} & Google &  54 \\
%     \cline{2-3}
%     {} & Yahoo & 32 \\
%     \hline
%     \end{tabular}
% \end{table}
\end{document}